\documentclass[aps,prd,onecolumn,showpacs,nofootinbib,groupedaddress, 11pt]{revtex4}  
\usepackage{graphicx,epstopdf,amsmath,amsfonts,amssymb,appendix,comment,bbold}
\usepackage{color,slashed,subfigure,setspace,footnote,multirow,longtable,braket}
\usepackage{epsfig,mathrsfs,latexsym,hyperref,color,url,etoolbox}
\usepackage{tikz-feynman}
\usepackage[letterpaper, portrait, margin=0.75in]{geometry}
\hypersetup{colorlinks,citecolor= nicegreen,linkcolor= nicered}
\definecolor{nicered}{rgb}{0.7,0.1,0.1}
\definecolor{nicegreen}{rgb}{0.1,0.5,0.1}
\usepackage{bbm}

\usetikzlibrary{external}
\immediate\write18{mkdir -p pgf-img}
\begin{document}

\singlespacing
\allowdisplaybreaks

{\hfill NUHEP-TH/18-05}

\title{Shining Light on the Mass Scale and Nature of Neutrinos with $e\gamma \to e\nu\overline{\nu}$}

\author{Jeffrey M. Berryman}
\affiliation{Center for Neutrino Physics, Department of Physics, Virginia Tech, Blacksburg, VA 24061, USA}
\author{Andr\'{e} de Gouv\^{e}a} 
\affiliation{Northwestern University, Department of Physics \& Astronomy, 2145 Sheridan Road, Evanston, IL 60208, USA}
\author{Kevin J. Kelly}
\affiliation{Northwestern University, Department of Physics \& Astronomy, 2145 Sheridan Road, Evanston, IL 60208, USA}
\author{Michael Schmitt}
\affiliation{Northwestern University, Department of Physics \& Astronomy, 2145 Sheridan Road, Evanston, IL 60208, USA}

\begin{abstract}
The discovery of neutrino oscillations invites many fundamental physics questions that have yet to be answered. Two of these questions are simple, easy to state, and essential: What are the values of the neutrino masses? Are neutrinos Majorana fermions? The reason we don't know the answer to those questions is that it is difficult to measure neutrino properties outside of the ultrarelativistic regime. We discuss the physics of $e\gamma\to e\nu\bar{\nu}$ near threshold, where one has access to nonrelativistic neutrinos and only nonrelativistic neutrinos. Near threshold, $e\gamma\to e\nu\bar{\nu}$ is a rich phenomenon and its cross section is sensitive to the individual values of the neutrino masses and the nature of the neutrinos. We show that if one could scan the threshold region, it would be simple to identify the mass of the lightest neutrino, the neutrino mass ordering, and whether the neutrinos are Majorana fermions. In practice, however, event rates are tiny and backgrounds are huge; the observation of $e\gamma\to e\nu\bar{\nu}$ in the sub-eV regime appears to be utterly inaccessible in the laboratory. Our results, nonetheless, effectively illustrate the discriminatory power of nonrelativistic neutrino observables. 
\end{abstract}

\pacs{PACS}

\maketitle

\setcounter{equation}{0}
\section{Introduction}
\label{sec:introduction}

The discovery of neutrino oscillations reveals that neutrinos do not behave as prescribed by the  Standard Model (SM).
The unambiguous feature of neutrino physics that has been unearthed from these oscillations is that at least two of the three known neutrinos have nonzero masses. Furthermore, these masses are orders of magnitude less than those of all other known fermions. In the last two decades, there has been great theoretical interest in trying to understand the mechanism by which neutrinos acquire these tiny masses. While numerous candidates have been proposed, none has emerged as the single most compelling explanation. Part of the difficulty is that, even two decades into this enterprise, there is still insufficient information on the neutrino masses themselves and the nature of the neutrinos is unknown. 

The magnitudes of the differences of the squares of the neutrino masses -- $\Delta m_{21}^2$ and $\Delta m_{31}^2$ -- have been measured at the $3$\% level \cite{Esteban:2016qun} by oscillation experiments, but the sign of the latter -- the neutrino mass hierarchy -- remains undetermined.\footnote{Modulo surprises, however, it is widely expected that the ambiguity in the neutrino mass hierarchy will be resolved with more data from the current generation of oscillation experiments.} There are physical observations sensitive to the overall scale of the neutrino masses, for instance, the shape of the endpoint of the spectrum for electrons produced in tritium beta decay depends on a particular combination of neutrino masses \cite{Kraus:2004zw,Aseev:2011dq,Osipowicz:2001sq,Arenz:2018kma}, and nonzero neutrino masses contribute to the observed anisotropies of the cosmic microwave background \cite{Ade:2015xua}. While the upper bounds derived from these phenomena are nontrivial and point to neutrino masses below the eV scale, a precise determination of the values of the neutrino masses is still lacking.

It is also unknown whether or not the neutrino and its antiparticle are distinct physical objects, i.e., whether neutrinos are Dirac or Majorana fermions. The answer to this question has profound impact on the mechanism responsible for nonzero neutrino masses. 
%
If neutrinos are Majorana fermions, there are additional avenues for pursuing the absolute neutrino masses. In particular, in a large class of models, the rates of lepton-number-violating processes will be governed by elements of the Majorana neutrino mass matrix. The most famous such process is neutrinoless double beta decay ($0\nu\beta\beta$), $(A, \, Z) \to (A, Z + 2) + 2e^-$ \cite{DellOro:2016tmg,Henning:2016fad}. Other possible lepton-number-violating processes include muon-to-positron conversion in nuclei \cite{Atre:2005eb,Berryman:2016slh,Geib:2016atx} as well as rare lepton \cite{Atre:2005eb,Kuno:2015tya,Quintero:2016iwi} and meson \cite{Atre:2005eb,Quintero:2016iwi} decays, but the bounds on these are significantly weaker than those derived from $0\nu\beta\beta$. Upper bounds on neutrino masses derived from $0\nu\beta\beta$ are competitive with the bounds from $\beta$-decay and cosmology but are moot if neutrinos are Dirac fermions.

In this paper, we discuss another process in which neutrino masses can be important: $e^- \gamma \to e^- \nu \overline{\nu}$, which we refer to as stimulated $\nu \overline{\nu}$ emission\footnote{While we will be interested in this process for both Dirac- and Majorana-type final state neutrinos, we will use Dirac-type language and refer to the process as $e\gamma\to e\nu\overline{\nu}$.}. Specifically, we consider this process when a sub-eV photon impinges on an electron at rest; we probe the threshold region, in which the final state neutrinos can be nonrelativistic. This is a SM process that, as far as we can tell, has not been previously discussed in the literature, and with good reason: its cross section is orders of magnitude below what could have been detected in any laboratory experiment to date. However, motivated by the landmark detection of coherent elastic neutrino-nucleus scattering at the COHERENT experiment \cite{Akimov:2017ade} more than four decades after its prediction \cite{Freedman:1973yd}, we ask what we would learn about neutrino properties if  stimulated $\nu \overline{\nu}$ emission were measured in a terrestrial setting and discuss the challenges associated with performing such a measurement. Additionally, we investigate the extent to which physics beyond the Standard Model can affect this process; since stimulated $\nu \overline{\nu}$ emission occurs with weak-interaction strength, it may be possible for stronger-than-weak physics to manifest itself in a nontrivial way.

The rest of this paper is organized as follows. In Sec.~\ref{sec:SMxsec}, we study stimulated $\nu \overline{\nu}$ emission in the Standard Model, for both Dirac and Majorana neutrinos, as well as either a normal or inverted hierarchy. In Sec.~\ref{sec:Backgrounds}, we consider event rates and the size of SM backgrounds to any potential search for this process. In Sec.~\ref{sec:NewPhysics}, we assess the impact of several new-physics scenarios on this process, focusing on (a) the introduction of a fourth, sterile neutrino, (b) a large magnetic or electric dipole moment induced by heavy new physics, and (c) the effects of a light ($\sim$MeV-scale) dark photon of gauged $B-L$. In Sec.~\ref{sec:Conclusions}, we offer some concluding thoughts.

\setcounter{equation}{0}
\section{Stimulated $\nu \overline{\nu}$ Emission}
\label{sec:SMxsec}

In the Standard Model, the process $e^- \gamma \to e^- \nu_i \overline{\nu}_j$ exists. If the electron is at rest, then the threshold energy of the incoming photon is
\begin{equation}
E_\gamma^\mathrm{thr.} = (m_i + m_j)\left(1 + \frac{m_i + m_j}{2m_e}\right),
\end{equation}
where $m_i$ is the mass of $\nu_i$, $m_j$ is the mass of $\overline{\nu}_j$, and $m_e$ is the mass of the electron. Regardless of the nature of the neutrinos (Majorana or Dirac fermions) or the ordering of their masses, this threshold behavior, at which only the lightest mass eigenstate can be produced, is a clear signal of the overall neutrino mass scale. As the photon energy increases, additional combinations of mass eigenstates become kinematically accessible, until enough energy exists to produce all states. When the photon energy is well above this final threshold, the distinction between mass-eigenstate and flavor-eigenstate final states becomes negligible.

Since the energies we will consider are far below the scale of electroweak physics, amplitudes will be calculated using the effective four-point $\ell\ell\nu\nu$ interaction obtained by integrating out the $W$ and $Z$ bosons; see, for instance, Ref.~\cite{Erler:2013xha} for details. The Lagrangian can be written
\begin{align}
\label{eq:WeakLagrangian}
\mathcal{L} & \supset \mathcal{L}_{\rm CC} + \mathcal{L}_{\rm NC}, \nonumber \\
\mathcal{L}_{\rm CC} & = -2\sqrt{2} G_F \left( \overline{\ell}_\alpha \gamma^\mu P_L \nu_\alpha \right) \left( \overline{\nu}_\beta \gamma_\mu P_L \ell_\beta \right), \\
\mathcal{L}_{\rm NC} & = \sqrt{2} G_F \left( \overline{\nu}_\alpha \gamma^\mu P_L \nu_\alpha \right) \left[ \overline{\ell}_\beta \gamma_\mu \left( \frac{(1-4s^2_W)\mathbbm{1} - \gamma_5}{2} \right) \ell_\beta \right], \nonumber
\end{align}
where $\mathcal{L}_{\rm CC}$ and $\mathcal{L}_{\rm NC}$ are the relevant charged- and neutral-current pieces of the Lagrangian, respectively, $G_F$ is the Fermi constant, $s_W$ is the sine of the Weinberg angle and $\alpha$, $\beta$ = $e, \, \mu, \, \tau$ are flavor indices. This Lagrangian can be rewritten in terms of the neutrino mass eigenstates by identifying
\begin{equation}
\label{eq:DefineU}
\nu_\alpha = U_{\alpha i} \nu_i,
\end{equation}
where $U$ is the unitary leptonic mixing matrix and $i = 1, \, 2, \, 3$ is a mass eigenstate index. This identity and a Fierz transformation allow us to recast the interactions as
\begin{equation}
\label{eq:interaction}
\mathcal{L}_{\rm CC} + \mathcal{L}_{\rm NC} = -\sqrt{2} G_F \left(\overline{\nu}_j \gamma^\mu P_L \nu_i \right) \left[ \overline{\ell}_\alpha \gamma_\mu \left( g_{V}^{\alpha\beta ij} \mathbbm{1} - g_{A}^{\alpha\beta ij} \gamma_5 \right) \ell_\beta \right],
\end{equation}
where we introduce the vector and axial couplings
\begin{equation}
\label{eq:OldGs}
g_V^{ \alpha \beta i j} = U_{\alpha i} U_{\beta j}^* - \frac{1}{2}\left(1 - 4\sin^2{\theta_W}\right) \delta_{ij} \delta_{\alpha \beta}, \qquad g_A^{\alpha \beta i j} = U_{\alpha i} U_{\beta j}^* - \frac{1}{2} \delta_{ij} \delta_{\alpha \beta}.
\end{equation}
Since the only charged leptons considered in this work are electrons, we will make the simplification $g_{V,A}^{ij} \equiv g_{V,A}^{eeij}$. 

The following diagrams are relevant to the evaluation of the amplitude:
\begin{equation}
\label{eq:FeynmanDiagrams}
i \mathcal{M} \approx 
\raisebox{-2.5em}{\begin{tikzpicture}
\begin{feynman}
\vertex (a1) {\(e^-\)};
\vertex[below=4em of a1](a2) {\(\gamma\)};
\vertex at ($(a1)!0.5!(a2) + (1cm, 0)$) (b1);
\node[small,blob] (c1) at ($(b1)+(1cm,0)$) {};
\vertex[right=1.5cm of c1] (d1) {\(e^-\)};
\vertex[below=2em of d1] (d2) {\(\overline{\nu}_j\)};
\vertex[above=2em of d1] (d3) {\(\nu_i\)};
\diagram*{ { [edges=fermion] (a1) -- (b1) -- (c1) -- (d1), (d2) -- (c1) -- (d3) }, (a2) -- [boson] (b1) };
\end{feynman}
\end{tikzpicture}}
+
\raisebox{-2.5em}{\begin{tikzpicture}
\begin{feynman}
\vertex (a1) {\(e^-\)};
\vertex[below=4em of a1](a2) {\(\gamma\)};
\vertex[right=1.5cm of a2] (b1);
\node[small,blob] (c1) at ($(a1)+(1.5cm,0)$) {};
\vertex[right=1cm of b1] (d1) {\(e^-\)};
\vertex[above=4em of d1] (d3) {\(\nu_i\)};
\vertex[above=2em of d1] (d2) {\(\overline{\nu}_j\)};
\diagram*{ { [edges=fermion] (a1) -- (c1) -- (b1) -- (d1), (d2) -- (c1) -- (d3) }, (a2) -- [boson] (b1) };
\end{feynman}
\end{tikzpicture}}.
\end{equation}
In Appendix~\ref{sec:Appendix}, we give analytic forms for the squared matrix element for our process of interest, assuming the neutrinos are either Dirac or Majorana fermions, in terms of the four-momenta of the incoming and outgoing states. In the case $i \neq j$, off-diagonal final states, the vector and axial couplings are equal. The $|\mathcal{M}|^2$ simplify significantly, then, all being proportional to $2|g_{V,A}^{ij }|^2 = 2|U_{e i}|^2|U_{e j}|^2$. Using experimental information on the magnitudes of the elements of the leptonic mixing matrix \cite{Esteban:2016qun}, the largest contribution to $|\mathcal{M}|^2$ comes from the channel $\{i, \, j\} = \{1, \, 2\}$ (or $ \{2, \, 1\}$). Ignoring effects due to available phase space, the $\{i, \, j\} = \{1, \, 3\}$ contribution is $\sim 8\%$ of that from $\{i, \, j\} = \{1, \, 2\}$, and the $\{i, \, j\} = \{2, \, 3\}$ channel is roughly $3-4\%$ of that from $\{i, \, j\} = \{1, \, 2\}$.

For diagonal final states, $i = j$, determining which channel dominates is less straightforward. There is no interference between the vector and axial-vector contributions for same-mass final-state neutrinos. Additionally, for $E_\gamma \ll m_e$, axial contributions dominate over vector ones. The $i=j$ final state that dominates for energies of interest will be the one with the largest $|g_A^{ii}|$ -- From Eq.~\ref{eq:OldGs}, we have $|g_A^{11}| = 0.18$, $|g_A^{22}| = 0.20$, and $|g_A^{33}| = 0.48$, so we expect ${i, j} = {3, 3}$ to dominate. For completeness, the vector couplings are $|g_V^{11}| = 0.62$, $|g_V^{22}| = 0.25$, and $|g_V^{33}| = 0.03$.

The cross section for stimulated $\nu\overline{\nu}$ emission is insensitive to the value of the $CP$-violating phase $\delta_{CP}$. When $i = j$, both the charged-current and neutral-current terms in $g_{V}^{ii}$ and $g_{A}^{ii}$ are real and independent of $\delta_{CP}$. When $i \neq j$, the neutral-current term vanishes and $g_{V}^{ij}$ and $g_{A}^{ij}$ are identical making their overall phase irrelevant. The analogous process involving muons or taus would, however, depend on $\delta_{CP}$, as changing $\delta_{CP}$ for fixed values of the mixing angles changes the magnitudes of $U_{\mu1}$, $U_{\mu2}$, $U_{\tau1}$ and $U_{\tau2}$. This is a consequence of the parameterization of the leptonic mixing matrix: fixing the mixing angles is sufficient to fix the magnitudes of the elements of its first row, whereas the other rows require additional input in the form of $\delta_{CP}$. Moreover, none of the $U_{ei}$ depend on $\theta_{23}$ in this parameterization; because one cannot determine if $\theta_{23}$ is nonzero from these elements, and because $\delta_{CP}$ is unphysical if any of the mixing angles vanish, the scattering cross section cannot depend on the value of $\delta_{CP}$. We emphasize that this dependence on the value of the $CP$-violating phase does not indicate that $CP$ symmetry is violated. Changing its value changes the magnitudes of $g_V^{ij}$ and $g_A^{ij}$, but it does not introduce a relative phase between them that could yield different cross sections for scattering involving positrons instead of electrons.\footnote{At higher order in the weak interactions, the cross sections for stimulated neutrino emission from positrons and electrons could be different and hence violate $CP$ invariance. We do not explore this issue here.}

\subsection{Normal vs. Inverted Mass Hierarchies}

The arguments made above regarding which final states dominate the overall cross section well above threshold are independent of the values of the neutrino masses $m_{1,2,3}$, and are a consequence purely of the leptonic mixing matrix. Because the elements of the mixing matrix are well known, we use their central values in our calculations, and explore how other properties of the neutrino mass spectrum influence this process.

The values of the neutrino mass splittings $\Delta m_{ij}^2 \equiv m_i^2 - m_j^2$ are also well constrained. What remains to be known, however, is the overall mass scale of the neutrinos. This is typically parameterized in terms of the lightest neutrino mass, which we designate $m_0$. We do not label it specifically $m_1$, given the possibility that the third mass eigenstate is the lightest. This latter scenario is known as the inverted mass hierarchy.

In the two mass hierarchies, normal (NH) and inverted (IH), for the same lightest neutrino mass $m_0$, different combinations of final state neutrinos become kinematically accessible for different photon energies. For instance, in the IH, the final state $\{i, \, j\} = \{3, \, 3\}$, which is the dominant diagonal contribution, is the first to become accessible, while it requires a significantly higher energy in the NH. The same is true for the final state $\{i, \, j\} = \{1, \, 2\}$ becoming available at lower energy for the NH than the IH.

\begin{figure}
\centering
\includegraphics[width=\linewidth]{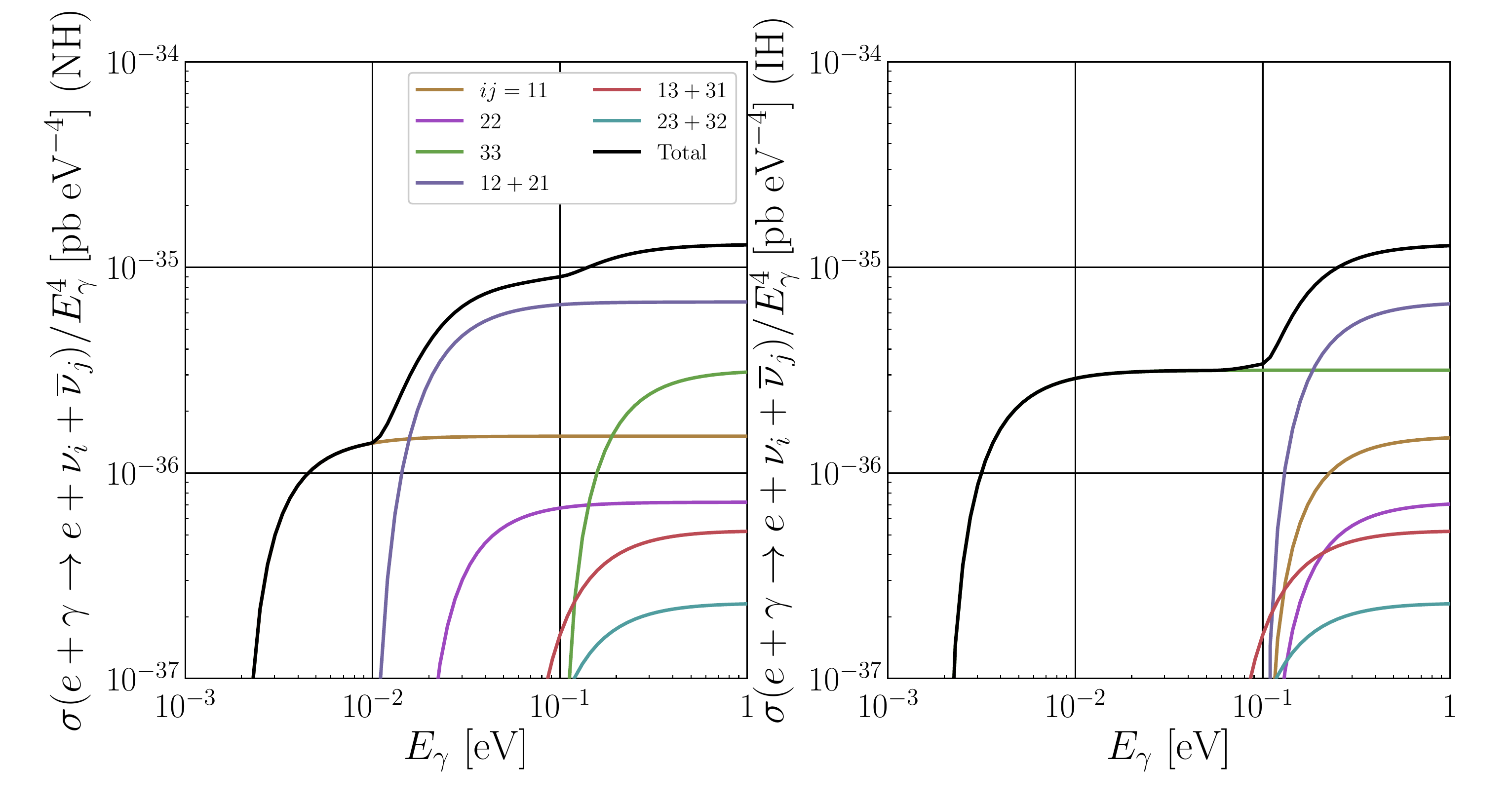}
\caption{Cross section of producing individual final states of neutrinos (colored) and total cross section (black) for a normal hierarchy (left) and inverted hierarchy (right) for $m_0 = 10^{-3}$ eV, as a function of the incoming photon energy $E_\gamma$. We divide the cross sections by $E_\gamma^4$, as the cross section grows with $E_\gamma^4$ well above threshold. Here, neutrinos are assumed to be Dirac particles.}
\label{fig:DifferentHierarchies}
\end{figure}

\begin{figure}
\centering
\includegraphics[width=\linewidth]{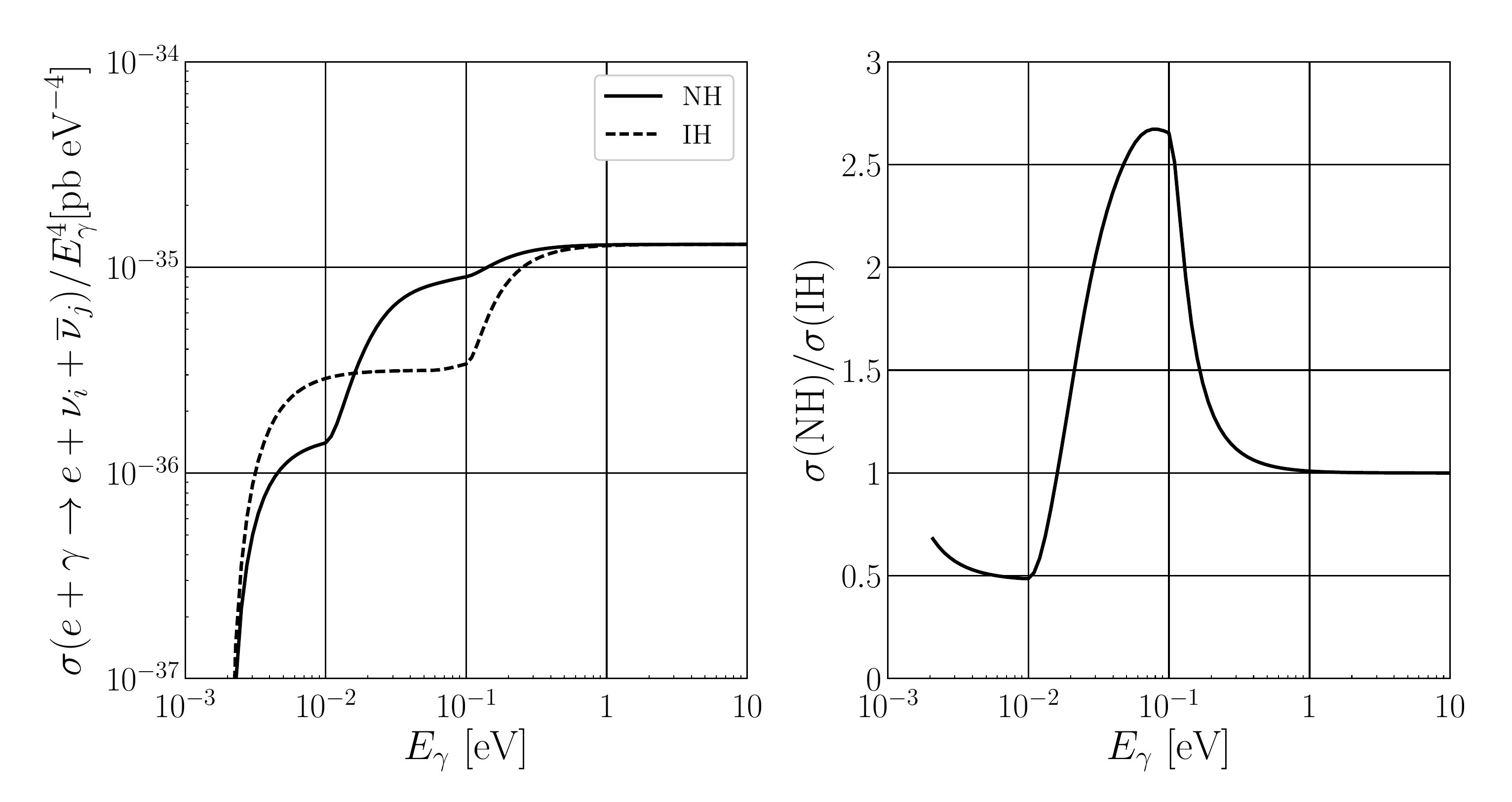}
\caption{Left: Total cross section for stimulated $\nu\overline{\nu}$ emission assuming a normal mass hierarchy (solid) and inverted mass hierarchy (dashed) with $m_0 = 10^{-3}$ eV. Right: Ratio between cross sections of normal and inverted mass hierarchies. Here, neutrinos are assumed to be Dirac particles.}
\label{fig:Ratio1D}
\end{figure}

For $m_0 = 10^{-3}$ eV and Dirac neutrinos, the cross section for each possible final state is depicted in Fig.~\ref{fig:DifferentHierarchies}.
For both hierarchies, the final states $\{i, \, j\} = \{1, \, 2\}$ (or $ \{2, \, 1\}$) and $\{i, \, j\} = \{3, \, 3\}$ are the most relevant at high energies. To highlight the difference between the total cross sections for the two different hierarchies, we show them together, along with the ratio between the two, in Fig.~\ref{fig:Ratio1D}.
This ratio deviates significantly from one in the region between the energies at which the $\{i, \, j\} = \{1, \, 2\}$ state becomes accessible for the two hierarchies. For $m_0 = 10^{-3}$ eV, this region is between $E_\gamma = 10^{-2}$ and $10^{-1}$ eV.

\begin{figure}
\centering
\includegraphics[width=0.75\linewidth]{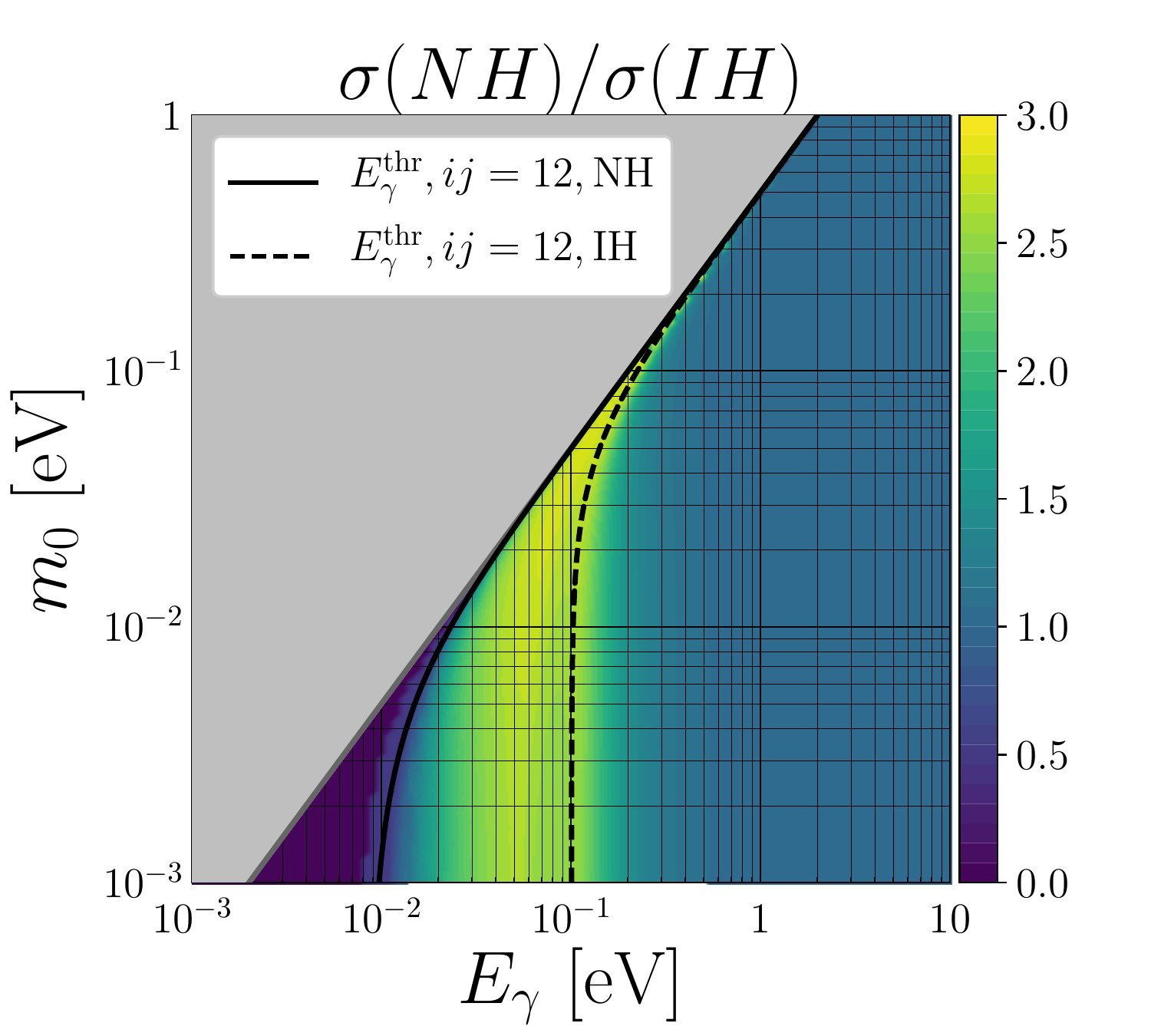}
\caption{Ratio of cross sections between the normal and inverted mass hierarchies with the same lightest neutrino mass $m_0$ as a function of the incident photon energy $E_\gamma$. Yellow regions correspond to regions in which the normal hierarchy cross section is larger, where purple regions correspond to regions in which the inverted hierarchy dominates. We also display the threshold energy for producing the dominant final state $\nu_1 \overline{\nu}_2 + \nu_2 \overline{\nu}_1$ for the normal (solid) and inverted (dashed) hierarchies. The gray region (top-left) has no kinematically accessible final states.}
\label{fig:RatioHierarchy}
\end{figure}

Extending to different values of $m_0$, the ratio between the NH and IH cross sections as a function of $E_\gamma$ and $m_0$ is depicted in Fig.~\ref{fig:RatioHierarchy}. Again, the region of interest for the largest deviations between these two cross sections is between the energies at which the $\{i, \, j\} = \{1, \, 2\}$ final state becomes accessible for the two hierarchies.

In summary, if one were to measure the threshold effect of $e\gamma \to e\nu\overline{\nu}$, then the lightest neutrino mass could be determined. A more detailed measurement of the excitation curve would reveal the mass hierarchy, because there are some qualitative features that distinguish between the two hierarchies near threshold. In Fig.~\ref{fig:Ratio1D}(left), there are three separate bumps in the cross section vs. $E_\gamma$ for the NH, from $\{i, \, j\} = \{1, \, 1\}$, $\{1, \, 2\}$, and $\{3, \, 3\}$ becoming kinematically accessible. For the IH, the only apparent bumps are from $\{i, \, j\} = \{3, \, 3\}$ and $\{i, \, j\} = \{1, \, 2\}$. Once the lightest mass is determined, the locations of these bumps are fixed by the magnitudes of $\Delta m_{21}^2$ and $\Delta m_{31}^2$.

\subsection{Dirac vs. Majorana Fermions}
The cross section for stimulated neutrino emission is different when one considers the final state neutrinos to be Dirac or Majorana fermions. This difference is proportional to $\sim (m_i m_j/s) \cos{(2\Delta \alpha_{ij})}$, where $\Delta \alpha_{ij} \equiv \alpha_j - \alpha_i$ is the difference between the Majorana phases associated to the mass eigenstates $\nu_i$ and $\nu_j$ and $s \simeq 2 m_e E_\gamma$ is the center-of-mass energy squared. As expected, the distinction between Majorana and Dirac neutrinos goes to zero as the photon energy increases. Therefore, there is an opportunity to discriminate between the two hypotheses near threshold. For $i = j$, because Dirac neutrinos have more available chirality states, the cross section is larger for Dirac neutrinos than for Majorana neutrinos. For $i \neq j$, the dominant  cross section depends on $\Delta \alpha_{ij}$.

We have computed the ratio between cross sections for Dirac and Majorana neutrinos close to threshold.  For more details, see Appendix~\ref{subsec:MajAnalytic}. The contribution of kinematic effects (e.g., thresholds) does not depend on the nature of the neutrino,\footnote{Several factors of $1/2$ and $2$ arise when comparing the Dirac and Majorana final state cross sections. These conspire in such a away that the total cross section well above threshold is the same for Majorana and Dirac neutrinos. For $i = j$, the Majorana final state cross section consists of two diagrams, however a factor of $1/2$ arises in the phase space integration because the final state particles are identical. For $i \neq j$, we compare the $\nu_i \nu_j$ Majorana final state to the $\nu_i \overline{\nu}_j + \nu_j \overline{\nu}_i$ Dirac final states, both of which have two distinct diagrams.} so all differences are properly captured by the squared matrix elements. If $i=j$, then $\Delta \alpha_{ij} = 0$, and 
\begin{equation}
\frac{\left\lvert \mathcal{M}_\mathrm{Dir.}\right\rvert^2}{\left\lvert \mathcal{M}_\mathrm{Maj.}\right\rvert^2} = \frac{3}{2} + \frac{\left\lvert g_V^{i i}\right\rvert^2}{2\left\lvert g_A^{i i}\right\rvert^2} + \left(1 + \frac{\left\lvert g_V^{i i}\right\rvert^2}{\left\lvert g_A^{i i}\right\rvert^2}\right) \frac{m_i}{m_e} + \mathcal{O}\left(\frac{m_i}{m_e}\right)^2.
\end{equation}
The value of this ratio depends strongly on the magnitude of $U_{ei}$. We estimate the matrix-element ratios to to be $7.9$, $2.1$, and $1.5$ for the final states $\{i, \, j\} = \{1, \, 1\}$, $\{2, \, 2\}$, and $\{3, \, 3\}$, respectively. On the other hand, if $i \neq j$,
\begin{equation}
\frac{\left\lvert \mathcal{M}_\mathrm{Dir.}\right\rvert^2}{\left\lvert \mathcal{M}_\mathrm{Maj.}\right\rvert^2} = \frac{1}{1 - \frac{1}{2}\cos{(2\Delta \alpha_{ij})}} + \frac{\cos{(2\Delta \alpha_{ij})}}{\left(\cos{(2\Delta \alpha_{ij})}-2\right)^2} \left(\frac{m_i + m_j}{m_e}\right) + \mathcal{O}\left(\frac{m_i + m_j}{m_e}\right)^2.
\end{equation}
Since the neutrino masses are small relative to that of the electron, and given that these ratios do not depend on the elements of the mixing matrix, they vary between $2/3$ and $2$, depending on $\Delta \alpha_{ij}$.

\begin{figure}
\centering
\includegraphics[width=\linewidth]{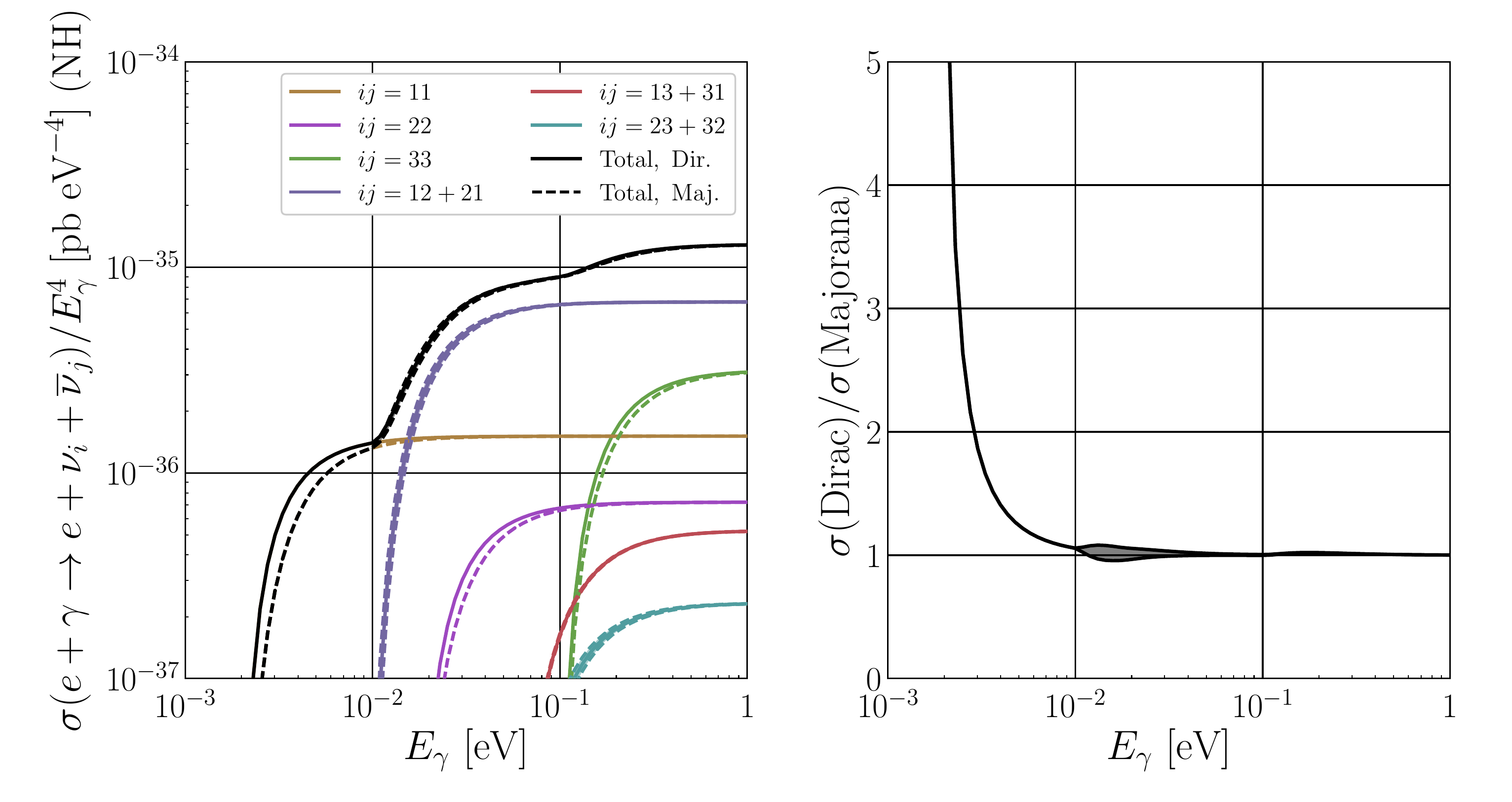}
\caption{Left: Cross sections for individual final states (colored) and total cross section for the process $e + \gamma \rightarrow e + \nu_i + \overline{\nu}_j$ assuming the neutrinos are Dirac particles (solid) and Majorana particles (dashed) with $m_1 = 10^{-3}$ eV and a normal mass hierarchy. For Majorana final states with $i \neq j$, the cross sections depend on Majorana phases and are depicted as (narrow) bands. Right: Ratio of Dirac to Majorana cross sections. The shaded region around $E_\gamma \sim 10^{-2}$ eV is the consequence of allowing for all possible values of the relative Majorana phase $\Delta \alpha_{12}$.}
\label{fig:MajDirNH}
\end{figure}

Figure~\ref{fig:MajDirNH}(left) depicts the cross sections of individual final states (colors) and the total cross section (black), assuming the neutrinos are Dirac (solid) and Majorana (dashed) fermions. For $i \neq j$, the cross sections for Majorana neutrinos depend on $\Delta \alpha_{ij}$, so the cross sections are bands (which are narrow on a logarithmic scale). Fig.~\ref{fig:MajDirNH}(right) depicts the ratio between the total cross sections for Dirac and Majorana final states. The dependence on the Majorana phases is more pronounced for values of the photon energy close to the region where the final state $\{i,j\}=\{1,2\}$ becomes kinematically accessible. Around $E_\gamma \sim 2 \times 10^{-2}$ eV, either Dirac or Majorana final states may have a larger cross section, depending on $\Delta \alpha_{12}$.

\begin{figure}
\centering
\includegraphics[width=\linewidth]{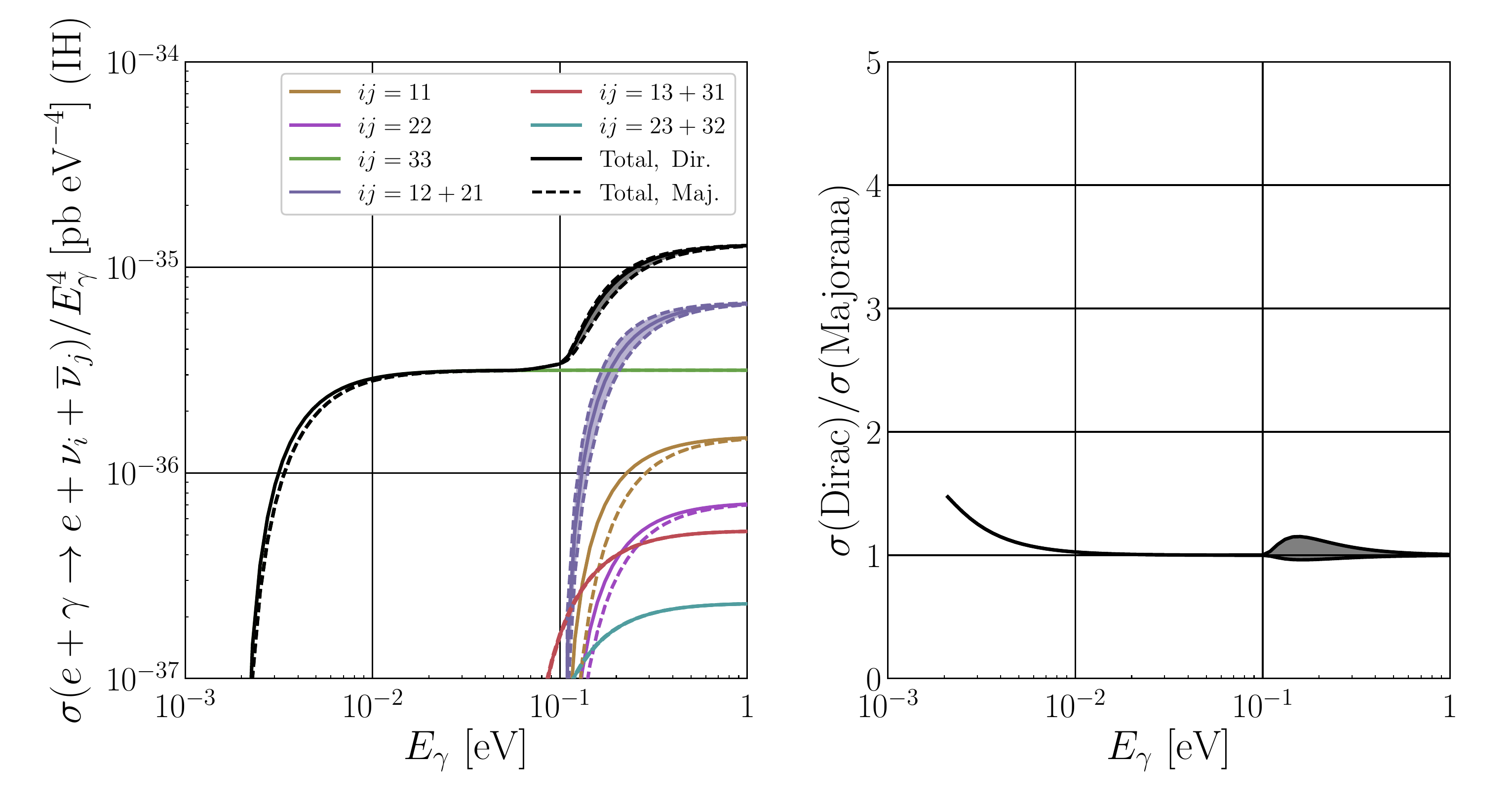}
\caption{Identical to Fig.~\ref{fig:MajDirNH}, however assuming $m_3 = 10^{-3}$ eV and an inverted mass hierarchy. Here  the ratio for Dirac vs. Majorana final states in the right panel is not as high as for the normal hierarchy, due to the difference in the vector and axial coupling constants for the $ij = 33$ final state.}
\label{fig:MajDirIH}
\end{figure}

Figure~\ref{fig:MajDirIH} depicts the same information as Fig.~\ref{fig:MajDirNH}, assuming an inverted mass hierarchy and $m_3 = 10^{-3}$ eV. In comparison with Fig.~\ref{fig:MajDirNH}, the Dirac/Majorana ratio does not become as large near threshold, since the axial coupling $|g_A^{33}|$ is larger than the vector coupling $|g_V^{33}|$. The spread in possible cross sections for the Majorana final state $\{i, \, j\}= \{1, \, 2\}$ is more pronounced here as well. The dependence of this ratio on the Majorana phases can be quite significant. For example, at $E_{\gamma} = 0.2$~eV in the IH, the cross section for the $\{i, \, j\}= \{1, \, 2\}$ final state varies between $6.1\times 10^{-39}$~pb and  $8.4\times 10^{-39}$~pb as one scans over all possible values of the Majorana phases.


\setcounter{footnote}{0}
\setcounter{equation}{0}
\section{Rates and Backgrounds}
\label{sec:Backgrounds}

In light of the cross section calculations of the previous section, we estimate the rate for stimulated $\nu \overline{\nu}$ emission in an unrealistically optimistic experimental setup. We imagine a configuration in which a $1$ eV ($1240$ nm, near-infrared) laser with a power of $\sim 2$ W~\cite{Sabella:10} is directed at a target of electrons at rest with density $n_e \sim 10^{23}$ cm$^{-3}$ and length of $1$ m. With the results of Sec.~\ref{sec:SMxsec}, such an experiment could expect one signal event every $\sim 10^{20}$ years. This cross section grows as $E_\gamma^4$ for $E_\gamma < m_e$, so the event rate improves rapidly for energies significantly above threshold. However, moving high enough above threshold to achieve an appreciable rate limits the usefulness of making this measurement, given our interest in probing the nonrelativistic regime.

Because of the extremely low energies involved, the photon-electron collisions under study benefit from the absence of hadronic activity in the final state. Moreover, since higher-order weak corrections will be strongly suppressed relative to the leading-order contribution, estimating the backgrounds is an exercise in pure QED. The dominant background for stimulated $\nu \overline{\nu}$ production is Thomson scattering, $e^- \gamma \to e^- \gamma$, the total cross section for which, in the limit $E_\gamma \ll m_e$ with unpolarized initial particles, is well known to be \cite{Peskin:1995ev}
\begin{equation}
\label{eq:thomson}
\sigma_{\text{Thomson}} = \frac{8 \pi \alpha^2}{3 m_e^2} = 0.667 \text{ b}.
\end{equation}
Because this is a $2\to2$ process, one could exploit the correlations between the energies of the final state electron and photon and their directions to identify these background events and separate them from the signal. However, compare this to the signal cross section calculated in the previous section for photon energies $E_\gamma = 1$ eV, $\sigma \sim \mathcal{O}\left(10^{-47}\right)$ b. This casts into sharp relief the infeasibility of detecting stimulated $\nu \overline{\nu}$ emission in a terrestrial experiment: for instance, one would need to ensure that no more than one in $\mathcal{O}(10^{47}-10^{48})$ Thomson photons escapes detection in order to effectively identify this background, an objective that is, for all intents and purposes, impossible.

To compound the matter, it is insufficient to consider single Thomson scattering as a background; one must also consider multiple Thomson scattering.\footnote{We are referring to a single event in which an incident photon is split into multiple final state photons, not to a single photon Thomson scattering multiple times.} Multiple Thomson scattering has been previously discussed in, for instance, Refs.~\cite{Ram:1971pi,Lotstedt:2012zz,Lotstedt:2013uya}. One would naively expect that the cross section for $n$-photon emission would scale as $\sigma_n \sim \alpha^{n-1} \sigma_{\text{Thomson}}$; while this is true in the high-energy regime ($E_\gamma \gtrsim m_e$), up to logarithmic corrections \cite{Sudakov:1954sw,Yennie:1961ad}, in the nonrelativistic regime the cross section has the form \cite{GouldPaper, Lotstedt:2013uya}
\begin{equation}
\sigma_n = C_n \alpha^{n-1} \sigma_{\rm Thomson} \times \left( \frac{E_\gamma}{m_e} \right)^{2n}, \, n \ge 2,
\end{equation}
where $C_n$ is a numerical coefficient of $\sim \mathcal{O}(0.1 - 10)$. The factor $(E_\gamma/m_e)^{2n}$ is a consequence of the necessity for a cutoff in photon energy. The rate of emission of multiple photons is formally divergent unless an infrared (IR) regulator is introduced; in an experiment, the cross section is regulated by the finite photon energy detection threshold. Introducing this IR regulator means that one is not integrating over all of phase space in calculating the cross section, and an additional suppression appears relative to the naive expectation.

On one hand, this extra suppression means that the backgrounds are not as daunting a challenge as one might originally suspect; one does not have to contend with $\mathcal{O}(20)$-photon backgrounds in order to study stimulated $\nu\overline{\nu}$ emission near threshold. On the other, we find that, for $E_\gamma \sim \mathcal{O}\left(10^{-2} - 1\right)$ eV, the cross section for $n=4, \, 5$ (depending on the choice of IR cutoff) still utterly swamps the signal cross section. Moreover, while one can, in principle, calculate angular and energy correlations between the final state particles to attempt to characterize and reduce these backgrounds, (1) the computational power required for a detailed study quickly becomes prohibitive (though not impossible), and (2) practical constraints exist on how well these backgrounds can be measured. 

The bottom line is that a terrestrial experiment will almost certainly never be able to measure stimulated $\nu \overline{\nu}$ emission at threshold. While we still regard this process as interesting, we move forward with the mentality that this process is one of purely theoretical relevance.


\setcounter{footnote}{0}
\setcounter{equation}{0}
\section{New Physics Contributions}
\label{sec:NewPhysics}

We consider how the presence of new physics beyond neutrino masses and lepton mixing would modify the results discussed in Sec.~\ref{sec:SMxsec}. We focus on three new-physics scenarios: the existence of a sterile neutrino, nonstandard neutrino electromagnetic properties, and the coupling of the SM neutrino to a dark photon.

\subsection{Sterile Neutrinos}
\label{subsec:FourNu}

Here, we consider the existence of a sterile neutrino species associated to a fourth neutrino mass eigenstate with mass around or below the eV scale. This scenario is qualitatively similar to the three-neutrino case outlined in the previous section, but mixing with the fourth neutrino induces important changes in the $\ell\ell\nu\nu$ interaction in Eq.~\eqref{eq:interaction}. Here, we simply quote the new values of the vector and axial coupling constants (Eq.~\eqref{eq:OldGs}); for details, see Appendix \ref{sec:Appendix2}:
\begin{equation}
\label{eq:NewGs}
g_V^{ i j} = U_{e i} U_{e j}^* - \frac{1}{2}\left(1 - 4\sin^2{\theta_W}\right) \left(  \delta_{ij} - U_{si} U_{sj}^* \right), \qquad g_A^{ i j} = U_{e i} U_{e j}^* - \frac{1}{2} \left(  \delta_{ij} - U_{si} U_{sj}^* \right),
\end{equation}
where $U_{si}$ is the $i$th element of the fourth row of the $4\times4$ extension of the leptonic mixing matrix. The inclusion of a sterile neutrino has two effects on these couplings. First, changes to the $U_{\alpha i}$ change the charged- and neutral-current contributions to $g_{V,A}^{ij}$ in a way that may either increase or decrease the magnitudes of the couplings. Second, the neutral-current contribution is no longer strictly diagonal in the mass basis, stemming from the absence of a coupling between the $Z$ boson and the sterile neutrino; see Appendix \ref{sec:Appendix2} for more details. 

\begin{table}[t]
\centering\begin{tabular}{c||c|c|c|c|c|c||c|c|c|c}
& $|U_{e2}|^2$ & $|U_{e3}|^2$ &  $|U_{e4}|^2$ &  $|U_{s2}|^2$ &  $|U_{s3}|^2$ &  $|U_{s4}|^2$ & $m_1$ [eV] & $\Delta m_{21}^2$ [eV$^2$] & $\Delta m_{31}^2$ [eV$^2$] & $\Delta m_{41}^2$ [eV$^2$] \\
\hline \hline
$\nu$SM & 0.30 & 0.022 & --- & --- & --- & --- & $10^{-3}$ & $7.40 \times 10^{-5}$ & $2.49 \times 10^{-3}$ & --- \\
\hline
Case 1 & 0.30 & 0.022 & 0.04 & 0.026 & 0.33 & 0.65 & $10^{-3}$ & $7.40 \times 10^{-5}$ & $2.49 \times 10^{-3}$ & $1.0 \times 10^{-5}$ \\
\hline
Case 2 & 0.30 & 0.022 & 0.01 & 0.001 & 0.024 & 0.99 & $10^{-3}$ & $7.40 \times 10^{-5}$ & $2.49 \times 10^{-3}$ & $1.3$ \\
\hline
Case 3 & 0.30 & 0.022 & 0.04 & 0.026 & 0.33 & 0.65 & $10^{-3}$ & $7.40 \times 10^{-5}$ & $2.49 \times 10^{-3}$ & $1.3$ \\
\hline
\end{tabular}
\caption{The neutrino-sector parameters used in our four-neutrino analysis. The values of $|U_{e2}|^2$ and $|U_{e3}|^2$ are fixed by measurements of 3$\nu$ oscillation parameters in Ref.~\cite{Esteban:2016qun}. For Cases 1 and 2, the other matrix elements are taken to be consistent with Ref.~\cite{Dentler:2018sju} for two different values of $\Delta m_{41}^2$. For Case 3, we have taken the mixing matrix elements for Case 1 with the $\Delta m_{41}^2$ in Case 2. For comparison, we show the parameters used in the $\nu$SM (SM plus nonzero neutrino masses of the Dirac or Majorana-type) analysis of Sec.~\ref{sec:SMxsec}. All $CP$-violating phases have been set to 0, for simplicity. The elements $|U_{e1}|^2$ and $|U_{s1}|^2$ can be inferred from the unitary of $U$.}
\label{tab:params}
\end{table}

We present three scenarios for the existence of sterile neutrinos with oscillation parameters given in Table \ref{tab:params}.  The first two cases are consistent with current bounds on the existence of a fourth neutrino presented in Ref.~\cite{Dentler:2018sju} for either a light ($\Delta m_{41}^2 = 1.0 \times 10^{-5}$ eV$^2$) or heavy ($\Delta m^2_{41} = 1.3$ eV$^2$) fourth neutrino; the latter of these is consistent with the global best-fit point from $\nu_e$/$\overline{\nu}_e$ disappearance data in Ref.~\cite{Dentler:2018sju}. The third case is already excluded at high confidence by existing oscillation data, but we use it to demonstrate important features of the cross section and to check the consistency of our calculations. In the rest of this work, we assume that the neutrino masses follow a normal hierarchy (i.e., $\Delta m_{31}^2 > 0$) and that the sign of $\Delta m_{41}^2$ is positive.

\begin{figure}[!t]
\centering
\includegraphics[width=1.0\linewidth]{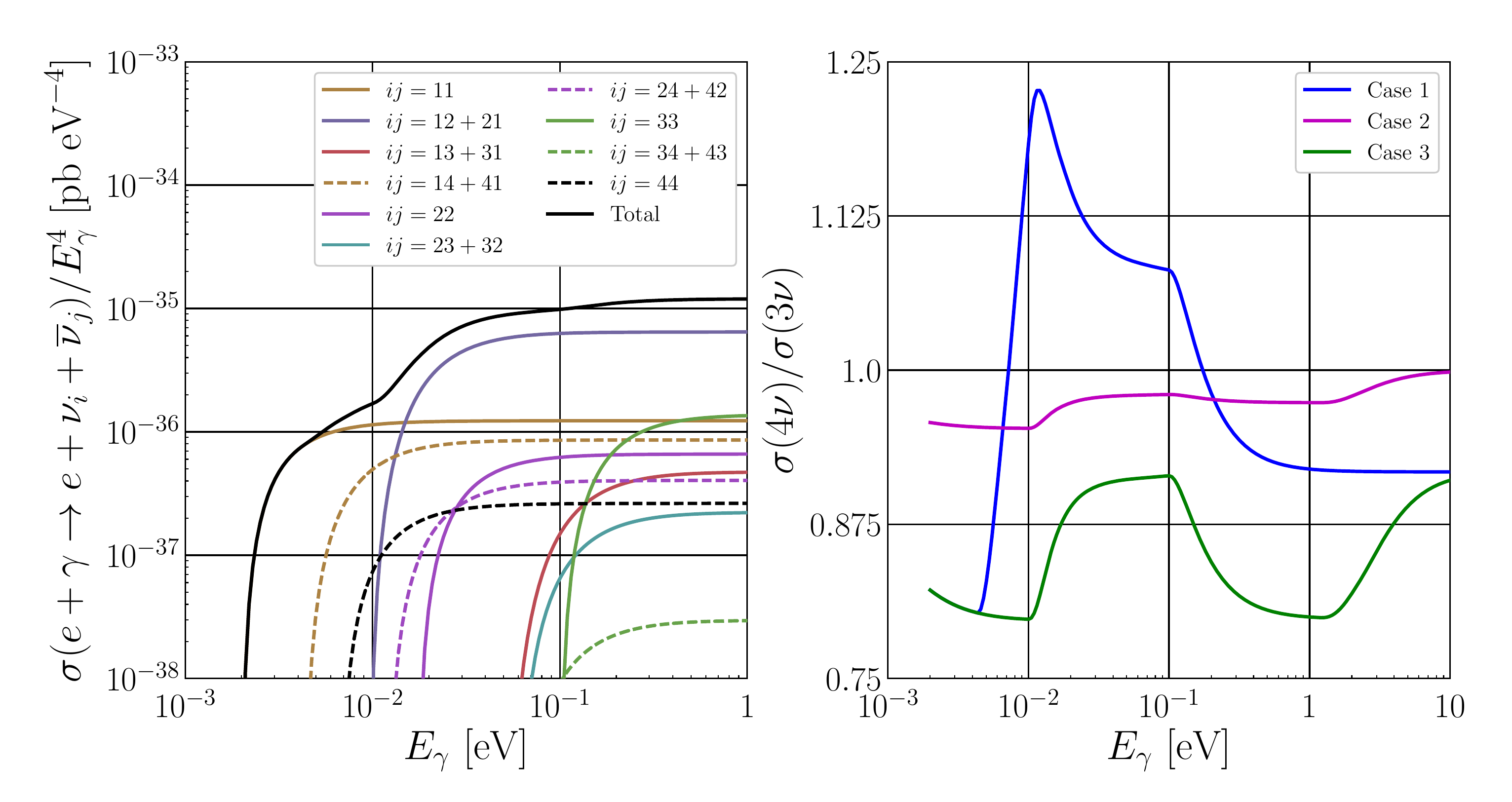}
\caption{Left: Cross sections for individual final states (colored) and total cross section for the process $e + \gamma \to e + \nu_i + \overline{\nu}_j$, assuming a fourth neutrino with parameters given for Case 1 in Table~\ref{tab:params}. Here, we assume neutrinos are Dirac particles with a normal mass hierarchy. Right: Ratio between cross sections assuming four neutrinos exist to the cross section assuming three neutrinos exist given parameters for Case 1 (blue), Case 2 (magenta), or Case 3 (green) from Table~\ref{tab:params}.}
\label{fig:SterileFig}
\end{figure}

The cross section for Case 1 is depicted in Fig.~\ref{fig:SterileFig} (left); we show the breakdown of the total cross section into components for only this case, in the interest of conserving space. In Fig.~\ref{fig:SterileFig} (right), we show how the total cross sections for all three cases in Table \ref{tab:params} compare to the three-neutrino cross section. In the high energy ($E_\gamma > 10$ eV) limit, Cases 1 and 3 are $\sim10$\% below the $\nu$SM (SM plus nonzero neutrino masses of the Dirac or Majorana-type) -- the relatively large mixing with the sterile neutrino results in a reduction of the charged-current contribution to the cross section -- while Case 2 is essentially consistent with the three-neutrino cross section. For lower energies, however, the differences can be quite sizable; the maximum excursion is $\sim 20\%$ for Cases 1 and 3, where the active-sterile mixing angles are relatively large, while this value is $\sim 5\%$ for Case 2. The convergence of Cases 1 and 3 at high energy indicates that our calculation is consistent: for $E_\gamma \gtrsim 10$ eV, the neutrino masses are irrelevant, so the overall cross section cannot distinguish between scenarios with different neutrino masses but the same mixing matrix. While these $\sim\mathcal{O}$(10\%) effects are not insignificant, we remind the reader that these effects are subdominant to other effects considered related to the physics of neutrinos -- the nature of the neutrinos, the mass hierarchy, etc. -- and could only be meaningfully observed after these other characteristics have been pinned down elsewhere.

Extending the mixing matrix introduces new sources of $CP$ violation, characterized by two new $CP$-violating phases. The placement of these new phases depends on the parameterization; we follow the conventions employed in Ref.~\cite{Berryman:2015nua}, where the three new phases, $\{ \eta_1, \, \eta_2, \, \eta_3 \}$, are the (negative of the) phases of $U_{e3}$, $U_{e4}$ and $U_{\mu 4}$, respectively. The presence of the $U_{si} U_{sj}^*$ contribution to the neural-current term in $g_{V,A}^{ij}$ spoils the independence of the stimulated $\nu\overline{\nu}$ emission cross section on these $CP$-violating phases that we encountered in Sec.~\ref{sec:SMxsec}. For $i = j$, these phases change the neutral-current contribution; for $i \neq j$, $g_{V}^{ij}$ and $g_{A}^{ij}$ are no longer equal and so may have a physically relevant relative phase.

To study the impact of these $CP$-violating phases on the cross sections, we compare the cross sections for Case 1 and Case 2 with all $CP$-violating phases set to zero against the same with each phase, in turn, set to $\pm \pi/2$. For Case 1, in which some of the mixing angles are sizable, varying the $CP$-violating phases can have a significant impact on the cross section. The largest deviation from the $CP$-conserving scenario is $\sim 12\%$ but is typically around $\sim 2-3\%$ in the threshold region. This effect is more modest for Case 2; due to the smallness of the mixing angles, the changes to the cross section are only as large as $\sim2\%$ and are mostly around the sub-percent level. Additionally, we have investigated the differences between $\eta_i = +\pi/2$ and $\eta_i = -\pi/2$ for each of the $\eta_i$ and find that the total cross section depends only on the magnitude of the phase. Despite the dependence of the cross section on these $CP$-violating phases, there is no $CP$ violation present in stimulated $\nu \overline{\nu}$ emission (at least at tree level).

\subsection{Neutrino Magnetic and Electric Dipole Moments}

If new physics exists but is heavier than available energies, then its associated degrees of freedom cannot be directly produced in experiments. However, it will still indirectly affect physical processes via modifications of leading-order couplings, or by introducing couplings that would not otherwise exist. This is typically described using the machinery of effective field theory \cite{Henning:2014wua,Brivio:2017vri}. In fact, this is precisely how we formulated our discussion in Sec.~\ref{sec:SMxsec}; the energies we consider are far below the electroweak gauge boson masses, but the effects of these are included via the Fermi operator. In this subsection, we will concern ourselves, in a model-independent way, with how heavy new physics may induce nonstandard electromagnetic properties for neutrinos. We can only hope to provide a cursory overview the subject; for in-depth discussions, see Refs.~\cite{Giunti:2014ixa,Giunti:2015gga,Balantekin:2018azf}.

The most general photon-neutrino-neutrino $\nu_i(k_i) \to \nu_j(k_j) + \gamma(q)$ vertex can be written as
\begin{equation}
\label{eq:vertex}
\feynmandiagram [baseline=(b.base), horizontal=a to b,small] {
a [particle=\(\nu_i\)] -- [fermion, momentum'=\(k_i\)] b [blob] -- [fermion, momentum'=\(k_j\)] c [particle=\(\nu_j\)],
b -- [boson, momentum'=\(q\)] d [particle=\(\gamma\)],
}; 
= i \left[ \left( \gamma_\mu - q_\mu \slashed{q}/q^2 \right) \left[ f^{ij}_Q(q^2) + q^2 f^{ij}_A(q^2) \gamma_5 \right] - i \sigma_{\mu\nu}q^\nu \left[ f^{ij}_M(q^2) + i f^{ij}_E(q^2) \gamma_5 \right] \right].
\end{equation}
The functions $f^{ij}_X(q^2)$, $X = Q,\, M,\, E, \, A$ are, respectively, the charge, magnetic dipole, electric dipole and anapole form factors which, in the limit $q^2 \to 0$, become the neutrino charge $q^{ij}_\nu$,\footnote{We rely on context to distinguish between the neutrino electric charge and $\nu$th component of the four-vector $q$.} magnetic dipole moment (MDM) $\mu^{ij}_\nu$, electric dipole moment (EDM) $\varepsilon^{ij}_\nu$, and anapole moment $a^{ij}_\nu$. If neutrinos are Dirac fermions, then the $f^{ij}_X$ are hermitian matrices in flavor space; if they are Majorana fermions, then the $f^{ij}_X$ are antisymmetric. We assume that, since the momenta are small relative to most energy scales in the process of interest, any dependence of the form factors on $q^2$ is negligible, allowing us to replace them with their respective moments.

The neutrino charge and anapole moments are severely constrained by existing data. A collection of bounds on $q_\nu$ are tabulated in Table IV of Ref.~\cite{Giunti:2014ixa}, but the authors of that reference derive $q_\nu = (-0.6 \pm 3.2) \times 10^{-21} e$ from the (non)neutrality of matter in Ref.~\cite{Bressi:2011yfa}. We find that neutrino charges of this order of magnitude make a negligible contribution to the stimulated $\nu\overline{\nu}$ emission cross section. Bounds on the neutrino anapole moments are derived from bounds on the neutrino charge radius, $\langle r_\nu^2 \rangle$; for exactly neutral neutrinos, we have $a_\nu = -\langle r^2_\nu \rangle/6$. Current limits on the neutrino charge radius are shown in Table V of Ref.~\cite{Giunti:2014ixa}; the strongest limits give $a_\nu \lesssim \mathcal{O}\left(10^{-32}\right)$ cm$^2$, an order of magnitude above their SM predictions \cite{Bernabeu:2000hf,Bernabeu:2002pd}. We have verified that anapole moments of this size are thoroughly subdominant to the SM contribution to the cross section of interest. We will therefore disregard neutrino charge and anapole moments and instead focus on MDMs and EDMs.

Some experimental limits on neutrino MDMs are tabulated in Table III of Ref.~\cite{Giunti:2014ixa}; more recent limits can be found in, for instance, Refs.~\cite{Borexino:2017fbd,Kosmas:2017tsq}. The strongest bounds from terrestrial experiments are $\mu_\nu \lesssim \mathcal{O}(10^{-11}) \mu_B$, where $\mu_B \, ( = 2.96 \times 10^{-7} \text{eV}^{-1})$ is the Bohr magneton \cite{Patrignani:2016xqp}; astrophysical limits can be as strong as $\mu_\nu \lesssim \mathcal{O}(10^{-12}) \mu_B$, but we will take $\mu_\nu = 10^{-11} \mu_B$ as our benchmark in this analysis. Moreover, every neutrino ever detected has been ultrarelativistic; in this regime, the vertex in Eq.~\eqref{eq:vertex} can be taken to be $\sigma_{\mu\nu}q^\nu (\mu_\nu - i \varepsilon_\nu)$, so experiments have only been able to constrain the quantity $ |\mu_\nu - i \varepsilon_\nu |$. As such, current bounds on $|\mu_\nu^{ij}|$ also apply to $|\varepsilon_\nu^{ij}|$. Refs.~\cite{Fujikawa:1980yx,Pal:1981rm,Shrock:1982sc} have calculated the neutrino MDMs and EDMs by minimally extending the SM with right-handed neutrinos -- making neutrinos Dirac particles -- and find
\begin{equation}
\label{eq:moments}
\left. \begin{array}{c} \mu_\nu^{ij} \\ i \varepsilon_\nu^{ij} \end{array} \right\} = \frac{3 e G_F}{16 \sqrt{2} \pi^2} (m_i \pm m_j) \times \left( \delta_{ij} - \frac{1}{2} \sum_{\ell = e, \, \mu , \, \tau} U^*_{\ell i} U_{\ell j} \frac{m_\ell^2}{M_W^2} \right) + \mathcal{O}\left(\frac{m_\ell^4}{M_W^4}\right),
\end{equation}
where the $+$ applies for $\mu_\nu^{ij}$  and the $-$ applies for $\varepsilon_\nu^{ij}$. Numerically, this expression becomes
\begin{align}
\mu_\nu^{ii} & \approx 3.2 \times 10^{-19} \mu_B \left( \frac{m_i}{\text{eV}} \right), \\
\varepsilon^{ii}_\nu & = 0,
\end{align}
for the diagonal elements, whereas for the off-diagonal elements, we find
\begin{equation}
\left. \begin{array}{c} \mu_\nu^{ij} \\ i \varepsilon_\nu^{ij} \end{array} \right\} = -3.9 \times 10^{-23} \mu_B \left( \frac{m_i \pm m_j}{\text{eV}} \right) \times \sum_{\ell = e, \, \mu , \, \tau} U^*_{\ell i} U_{\ell j} \frac{m_\ell^2}{m_\tau^2}. \qquad \qquad (i \neq j)
\end{equation}
Experimental bounds on $\mu_\nu^{ij}$ and $\varepsilon_\nu^{ij}$ are many orders of magnitude above the predicted values from these scenarios, leaving plenty of opportunity for this new physics to be discovered in, for instance, stimulated $\nu\bar{\nu}$ emission.

Generic theoretical arguments for the expected sizes of Majorana neutrino MDMs and EDMs have also been employed in the literature \cite{Shrock:1982sc} (see also the references in Sec.~IV~B of Ref.~\cite{Giunti:2014ixa}), but are more model dependent and we will not discuss them here explicitly. We will remark, however, that MDMs and EDMs can be large in several classes of models -- potentially many orders of magnitude larger than the predictions discussed above -- and constitute precisely the kind of new physics that motivates these searches.

\begin{figure}
\centering
\includegraphics[width=0.6\linewidth]{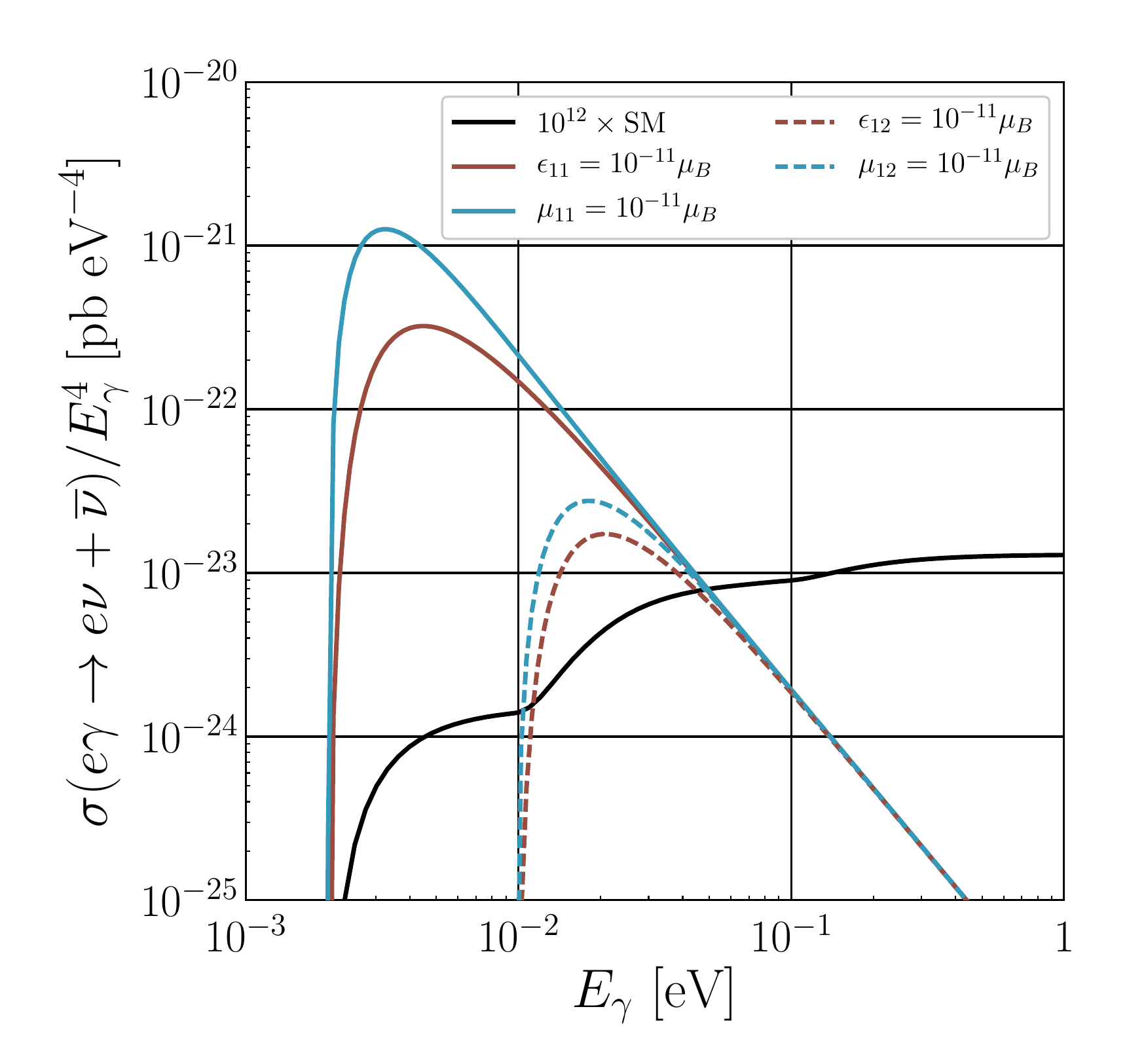}
\caption{Total cross section for an electric dipole moment $\epsilon_{ij} = 10^{-11} \mu_B$ (red) or magnetic dipole moment $\mu_{ij} = 10^{-11} \mu_B$ (blue), for $ij = 11$ (solid) and $ij = 12$ (dashed). Here, we also display the total cross section discussed in Section~\ref{sec:SMxsec}, increased by a factor of $10^{12}$ for comparison. Here we assume that neutrinos are Dirac fermions and the mass hierarchy is normal.}
\label{fig:MagElecMoments}
\end{figure}

Figure \ref{fig:MagElecMoments} depicts $e\gamma\to e\nu_i\overline{\nu}_j$ cross sections for either a nonzero Dirac MDM (blue lines) or a nonzero Dirac EDM (red lines) for the $\nu_1 \overline{\nu}_1$ final state (solid lines) and the $\nu_1 \overline{\nu}_2 + \nu_2 \overline{\nu}_1$ final state (dashed lines). In both cases, we have chosen $\mu_\nu^{ij}=\epsilon_\nu^{ij} =10^{-11} \mu_B$ for all $i,j$ and have only kept diagrams that are first order in the neutrino MDM/EDM, to wit,
\begin{equation}
i \mathcal{M} \approx 
\raisebox{-2.5em}{\begin{tikzpicture}
\begin{feynman}
\vertex (a1) {\(e^-\)};
\vertex[below=4em of a1](a2) {\(\gamma\)};
\vertex at ($(a1)!0.5!(a2) + (1cm, 0)$) (b1);
\vertex[right=1cm of b1] (c1);
\node[small,blob] (c2) at ($(c1)+(0.75cm,1em)$) {};
\vertex[right=4cm of a2] (d1) {\(e^-\)};
\vertex[above=2em of d1] (d2) {\(\overline{\nu}_j\)};
\vertex[above=4em of d1] (d3) {\(\nu_i\)};
\diagram*{ { [edges=fermion] (a1) -- (b1) -- (c1) -- (d1), (d2) -- (c2) -- (d3) }, (a2) -- [boson] (b1) , (c1) -- [boson, edge label = \(\gamma \)] (c2)};
\end{feynman}
\end{tikzpicture}}
+
\raisebox{-2.5em}{\begin{tikzpicture}
\begin{feynman}
\vertex (a1) {\(e^-\)};
\vertex[below=4em of a1](a2) {\(\gamma\)};
\vertex[right=1.5cm of a2] (b1);
\vertex[right=1.5cm of a1] (c1);
\node[small,blob] (c2) at ($(c1)+(1cm,0)$) {};
\vertex[right=2cm of b1] (d1) {\(e^-\)};
\vertex[above=4em of d1] (d3) {\(\nu_i\)};
\vertex[above=2em of d1] (d2) {\(\overline{\nu}_j\)};
\diagram*{ { [edges=fermion] (a1) -- (c1) -- (b1) -- (d1), (d2) -- (c2) -- (d3) }, (a2) -- [boson] (b1), (c1) -- [boson,edge label'=\(\gamma\)] (c2) };
\end{feynman}
\end{tikzpicture}}.
\end{equation}
The SM contribution discussed is also shown in black, multiplied by $10^{12}$ for comparison. There are a couple of factors that govern the relative sizes of weak and MDM/EDM cross sections. The first is that the MDM/EDM considered here is significantly larger than the predictions of Eq.~\eqref{eq:moments}. One would expect that a MDM/EDM of SM strength should appear as a one-loop weak correction to the tree-level $\nu$SM cross section and is hence completely negligible even close to $\nu\overline{\nu}$ threshold, as one can infer from Fig.~\ref{fig:MagElecMoments}.
The second factor is that since neutrinos are light, the cross section benefits from a nearly-soft, nearly-collinear enhancement from the virtual photon propagator; this overcomes the energy dependence of the vertex, resulting in a significant enhancement of the differential cross section. There is also an obvious difference in scaling of the cross section with energy -- $\sigma \sim E_\gamma^2$ here, compared with $\sigma \sim E_\gamma^4$, as before -- meaning the magnetic moment contribution would dominate the cross section for energies below $\sim (1 \text{ MeV}) \times \left( \dfrac{\mu_\nu \text{ or } \varepsilon_\nu}{10^{-11} \mu_B} \right)$.

At high energies, the MDM- and EDM-induced cross sections in Fig.~\ref{fig:MagElecMoments} converge. In the ultrarelativistic limit, the different-chirality final states decouple and the cross section only depends on $|\mu_\nu^{ij} \pm i \varepsilon^{ij}_\nu|$, making it impossible to disentangle the effects of a nonzero MDM and a nonzero EDM. However, a difference between nonzero MDMs and nonzero EDMs emerges at low energies: the $\gamma_5$ that accompanies $\varepsilon_\nu^{ij}$ (see Eq.~\eqref{eq:vertex}) causes the different-chirality final states to destructively interfere with one another, whereas the implicit $\mathbbm{1}$ that accompanies $\mu^{ij}_\nu$ allows these contributions to interfere constructively. If one could detect this process when the final state neutrinos are not ultrarelativistic, then one should be able to disentangle the effects of a nonzero MDM and nonzero EDM by measuring the shape of the cross section in this regime.

Dirac and Majorana dipole moments are qualitatively similar but with two important distinctions. (1) The Majorana dipole moment matrices must be antisymmetric, i.e., only transition moments exist, whereas the Dirac moment matrices are only constrained to be Hermitian. The absence of diagonal final states would be a strong (though not necessarily ironclad) indication of the Majorana nature of neutrinos. (2) For Majorana magnetic dipole moments, we must make the replacement $|\mu_\nu^{ij}|^2 \to 4|\mu_\nu^{ij}|^2$, and similarly for electric dipole moments, in the calculations above. As with Majorana masses, this change in dependency arises from interference between the $\nu_i \overline{\nu}_j$ and $\nu_j \overline{\nu}_i$ final states, but unlike the former scenario, this interference does not vanish at high energies.

\subsection{Gauged $U(1)_{B-L}$ Interactions}

In this subsection, we discuss the consequences of one particular model of new physics: the gauging of baryon-number-minus-lepton-number symmetry, $U(1)_{B-L}$. While this particular model faces strong constraints, we use it to represent how similar (though more complicated) models may result in cross sections that can rival the SM prediction in magnitude.

We introduce a new gauge boson $X$, which we refer to as the dark photon, that couples to the SM fermions proportionally to their $B-L$ charge. The relevant part of the new-physics Lagrangian is
\begin{equation}
\label{eq:BmL}
\mathcal{L}_{\text{new}} \supset -g_X X^\mu \cdot \left( \overline{e} \gamma_\mu e + \overline{\nu} \gamma_\mu \nu \right) + \frac{1}{2} M_X^2 X^\mu X_\mu - \frac{1}{4} X^{\mu \nu} X_{\mu \nu},
\end{equation}
where $g_X$ is the new gauge coupling and $M_X$ is the dark photon mass. We have only shown the couplings of $X$ to leptons, though it must also couple to quarks, and we ignore potential kinetic mixing between the $X$ and the SM photon. In order for $B-L$ to be free of anomalies, one must introduce a right-handed neutrino. Consequently, we assume neutrinos to be Dirac particles. Again ignoring contributions from the weak interactions, the amplitude for $\gamma + e^- \to e^- + \nu + \overline{\nu}$ can be calculated from the following diagrams:
\begin{equation}
i \mathcal{M} \approx 
\raisebox{-2.5em}{\begin{tikzpicture}
\begin{feynman}
\vertex (a1) {\(e^-\)};
\vertex[below=4em of a1](a2) {\(\gamma\)};
\vertex at ($(a1)!0.5!(a2) + (1cm, 0)$) (b1);
\vertex[right=1cm of b1] (c1);
\vertex at ($(c1)+(0.75cm,1em)$) (c2);
\vertex[right=4cm of a2] (d1) {\(e^-\)};
\vertex[above=2em of d1] (d2) {\(\overline{\nu}_j\)};
\vertex[above=4em of d1] (d3) {\(\nu_i\)};
\diagram*{ { [edges=fermion] (a1) -- (b1) -- (c1) -- (d1), (d2) -- (c2) -- (d3) }, (a2) -- [boson] (b1) , (c1) -- [boson, edge label = \(X \)] (c2)};
\end{feynman}
\end{tikzpicture}}
+
\raisebox{-2.5em}{\begin{tikzpicture}
\begin{feynman}
\vertex (a1) {\(e^-\)};
\vertex[below=4em of a1](a2) {\(\gamma\)};
\vertex[right=1.5cm of a2] (b1);
\vertex[right=1.5cm of a1] (c1);
\vertex (c2) at ($(c1)+(1cm,0)$);
\vertex[right=2cm of b1] (d1) {\(e^-\)};
\vertex[above=4em of d1] (d3) {\(\nu_i\)};
\vertex[above=2em of d1] (d2) {\(\overline{\nu}_j\)};
\diagram*{ { [edges=fermion] (a1) -- (c1) -- (b1) -- (d1), (d2) -- (c2) -- (d3) }, (a2) -- [boson] (b1), (c1) -- [boson,edge label'=\(X\)] (c2) };
\end{feynman}
\end{tikzpicture}}.
\end{equation}
As is the case of the SM neutral-current interactions, the final state neutrino and antineutrino must be in identical mass eigenstates. 

Limits on the $U(1)_{B-L}$ dark photon mass and coupling can be found in, for instance, Refs.~\cite{Harnik:2012ni,Cerdeno:2016sfi}. Null searches for a fifth, long-range force constrain the dark photon mass to be $\gtrsim \mathcal{O}$(100 eV); for sub-eV incident photon energies, the dark photon will always be off-shell, and we can treat the new interaction in the framework of effective field theory, similar to what we did previously for the weak interactions. We incorporate the effects of this dark photon at low energies by modifying the vector coupling $g_V^{ij}$:
\begin{equation}
g_V^{ij} \to g_V^{ij} + \delta_{ij} \frac{g_X^2}{\sqrt{2} G_F M_X^2} \equiv g_V^{ij} + \delta_{ij} \frac{G_X}{\sqrt{2} G_F},
\end{equation}
where $G_X$ is the Fermi-like coupling of gauged $U(1)_{B-L}$. There is a sliver of allowed parameter space in Refs.~\cite{Harnik:2012ni,Cerdeno:2016sfi} that allows for a $U(1)_{B-L}$ dark photon with $g_X = 10^{-6}$ and $M_X = 1$ MeV, corresponding to $G_X/G_F \lesssim 0.1$. As such, a dark photon in this region of parameter space would only contribute subdominantly to the weak interactions.

There are available regions of parameter space, particularly at higher $M_X$, where the coupling can be quite large: for instance, the point $g_X = 10^{-4}$, $M_X = 1$ GeV is allowed by current experiments. However, the increase in the coupling constant in the numerator of $G_X$ is never enough to outstrip the growth in its denominator, so this high-mass region cannot yield larger cross sections than those of the previous paragraph. For lower values of $M_X$, however, there exists a region of parameter space -- $g_X \sim \mathcal{O}(10^{-7})$ and $M_X \sim \mathcal{O}$(1 keV) -- where such a dark photon may be safe from solar cooling constraints; see Refs.~\cite{Harnik:2012ni,Cerdeno:2016sfi} for details. In this region of parameter space, not only do we have $G_X/G_F \sim 10^3$, resulting in new physics dominating over the SM, but if the energy of the incident photon were high enough ($E_\gamma \gtrsim M_X$), then the effective-operator approach would break down and there would be an additional enhancement to the cross section from the production of on-shell $X$. While it is intriguing that this process may be sensitive to such a strong, new force,\footnote{Because this interaction is diagonal and flavor universal, it is unconstrained by searches for nonstandard neutrino interactions in oscillation experiments.} our interest in this work is around the threshold region, so we do not consider this scenario further. We conclude this section by noting that the contribution of the dark photon of $U(1)_{B-L}$ to the neutrino MDM and EDM is highly suppressed, and does not lead to enhanced contributions of the type discussed in the previous subsection.

\setcounter{footnote}{0}
\setcounter{equation}{0}
\section{Discussion and Conclusions}
\label{sec:Conclusions}

The discovery of neutrino oscillations reveals that neutrinos have nonzero masses and leptons mix. It also invites many fundamental physics questions that have the potential to qualitatively change our understanding of particle physics. Two of these questions are simple, easy to state, and essential: what are, even roughly, the values of the neutrino masses -- we only have information on the neutrino mass-squared differences -- and are neutrinos Majorana fermions? The reason we don't know the answers to those questions is that neutrino masses are very small compared to the typical laboratory
neutrino energies in experiments.

We discussed a process that involves nonrelativistic neutrinos: $e\gamma\to e\nu\bar{\nu}$. When the neutrinos are nonrelativistic, $e\gamma\to e\nu\bar{\nu}$ is a rich phenomenon and the cross section is sensitive to the individual values of the neutrino masses and the Dirac or Majorana nature of the neutrinos. If one could scan the threshold region, then it would be simple to identify the mass of the lightest neutrino, the neutrino mass ordering, and whether the neutrinos are Dirac or Majorana fermions. As we remarked on several occasions -- see, for example, Figs.~\ref{fig:Ratio1D} and \ref{fig:MajDirNH} -- the cross section can change by up to $\mathcal{O}(100\%)$ depending on the answer to the individual question. The fact that we can distinguish Majorana neutrinos from Dirac neutrinos -- and measure the Majorana phases -- is intriguing and can be understood. At very low energies, lepton-number-violating phenomena are present and unsuppressed: what we would call $\nu\bar{\nu}$, $\nu\nu$, and  $\bar{\nu}\bar{\nu}$ final states are all present.

There are a few other processes that provide access to nonrelativistic neutrinos. The $2\to4$ scattering process $e^-e^{\pm}\to \ell^-\ell^{\prime\pm}\nu\bar{\nu}$ ($\ell,\ell'=e,\mu,\tau$) could also be studied at low energies and the outgoing charged leptons would contain enough information to measure the properties of the neutrinos, similar to what we discussed here. Nuclear $\beta$-decay also includes nonrelativistic neutrinos close to the end point of the electron energy spectrum. In principle, details around the endpoint of the spectrum contain information about the individual neutrino masses (see Ref.~\cite{Farzan:2001cj} and references therein). Nuclear $\beta$-decay energy spectra, however, cannot be used to distinguish Majorana from Dirac neutrinos, as there is only one neutrino in the final state. At threshold, $e\gamma\to e\nu\bar{\nu}$ is different in the sense that all neutrinos involved are nonrelativistic and the neutrinos are pair-produced; it is sensitive to both the neutral  and charged currents and there is a manifest distinction between Majorana and Dirac fermions. The cosmic neutrino background also serves as an intense, rich source of cold neutrinos that are mostly nonrelativistic. While they are yet to be observed, future precision measurements could access information about the neutrino masses and the nature of the neutrino \cite{Long:2014zva}.

Low-energy $e\gamma\to e\nu\bar{\nu}$ and the processes summarized above necessarily suffer from tiny cross sections because neutrino masses are very small and the weak interactions are aptly named at low energies. In Sec.~\ref{sec:Backgrounds}, we discussed event rates and backgrounds to $e\gamma\to e\nu\bar{\nu}$. In particular, multiple-photon backgrounds utterly overwhelm the $\nu \overline{\nu}$ signal, rendering a measurement of the latter virtually impossible in a terrestrial context. Nonetheless, $e\gamma\to e\nu\bar{\nu}$ processes should occur at nonnegligible rates in astrophysical systems. For instance, the flux of keV-scale (and below) neutrinos produced by several thermal production mechanisms in the Sun -- including stimulated $\nu \overline{\nu}$ emission -- has been calculated in Ref.~\cite{Vitagliano:2017odj}. That work indicates that the thermal neutrino flux would dominate the flux of neutrinos produced by proton-proton fusion at $\sim$ keV energies (see Fig.~18 therein). The spectrum of the solar neutrinos at sub-eV energies contains many of the features discussed in this work.  A detailed calculation of the role of threshold effects in $\nu \overline{\nu}$ production in astrophysical sources is beyond the scope of this work.

\section*{Acknowledgements}

Amplitudes have been evaluated using the Mathematica package FeynCalc \cite{Mertig:1990an,Shtabovenko:2016sxi}. Feynman diagrams have been drawn using the \LaTeX \, package TikZ-Feynman \cite{Ellis:2016jkw}. JMB is supported by DOE grant \#de-sc0018327 and acknowledges the support of the Colegio de F\'isica Fundamental e Interdiciplinaria de las Am\'ericas (COFI) Fellowship Program. AdG, KJK, and MS are supported in part by DOE grant \#de-sc0010143. KJK thanks the Fermilab Neutrino Physics Center for support during the work of this manuscript.


\appendix
\section{Standard Model Contributions}
\label{sec:Appendix}
Our process of interest is
\begin{equation}
\gamma (p_1) + e^- (p_2) \longrightarrow e^- (k_3) + \nu_i (k_4) + \overline{\nu}_j (k_5),
\end{equation}
which consists of the two diagrams in Eq.~\eqref{eq:FeynmanDiagrams}. We ignore a possible five-point diagram that arises in the effective theory from a photon coupling directly to the $W$ boson, whose corresponding amplitude is suppressed relative to the others by a factor $\sim E_\gamma^2 G_F$. The two matrix elements for this process are
\begin{align}
i\mathcal{M}_1  & = 
\raisebox{-4em}{\begin{tikzpicture}
\begin{feynman}
\vertex (a1) {\(e^-\)};
\vertex[below=7em of a1](a2) {\(\gamma\)};
\vertex at ($(a1)!0.5!(a2) + (1cm, 0)$) (b1);
\node[small,blob] (c1) at ($(b1)+(1cm,0)$) {};
\vertex[right=2cm of c1] (d1) {\(e^-\)};
\vertex at ($(d1)+(-0.75cm,3.5em)$) (d3) {\(\nu_i\)};
\vertex at ($(d1)+(-0.75cm,-3.5em)$)(d2) {\(\overline{\nu}_j\)};
\diagram*{ 
(a1) -- [fermion,momentum=\(p_2\)] (b1) -- [fermion] (c1) -- [fermion,momentum=\(k_3\)] (d1), 
(d2) -- [fermion,reversed momentum= \(k_5\)] (c1) -- [fermion,momentum=\(k_4\)] (d3) , 
(a2) -- [boson, momentum'=\(p_1\)] (b1) };
\end{feynman}
\end{tikzpicture}}
 \nonumber \\
& = \frac{e G_F}{\sqrt{2} (s - m_e^2)} \epsilon_\nu (p_1) \left[\overline{u}_{k_3} \gamma^\mu (g_V^{ i j} - g_A^{ i j}\gamma^5) (\slashed{p_1} + \slashed{p_2} + m_e) \gamma^\nu u_{p_2} \right] \left[ \overline{u}_{k_4} \gamma_\mu (1 - \gamma^5) v_{k_5} \right],\\
i\mathcal{M}_2 & = 
\raisebox{-2.5em}{\begin{tikzpicture}
\begin{feynman}
\vertex (a1) {\(e^-\)};
\vertex[below=4em of a1](a2) {\(\gamma\)};
\vertex[right=2cm of a2] (b1);
\node[small,blob] (c1) at ($(a1)+(2cm,0)$) {};
\vertex[right=1.75cm of b1] (d1) {\(e^-\)};
\vertex[above=2em of d1] (d2) {\(\overline{\nu}_j\)};
\vertex[above=6em of d1] (d3) {\(\nu_i\)};
\diagram*{ 
(a1) -- [fermion, momentum=\(p_2\)] (c1) -- [fermion] (b1) -- [fermion, momentum=\(k_3\)] (d1), 
(d2) -- [fermion, reversed momentum' = \(k_5\) ](c1) -- [fermion, momentum=\(k_4\)](d3),
 (a2) -- [boson, momentum=\(p_1\)] (b1) };
\end{feynman}
\end{tikzpicture}} \nonumber \\
 &= \frac{e G_F}{\sqrt{2} ((k_3 - p_1)^2 - m_e^2)} \epsilon_\nu (p_1) \left[ \overline{u}_{k_3} \gamma^\nu (\slashed{k_3} - \slashed{p_1} + m_e) \gamma^\mu (g_V^{ i j} - g_A^{ i j}\gamma^5) u_{p_2} \right] \left[ \overline{u}_{k_4} \gamma_\mu (1 - \gamma^5) v_{k_5} \right],
\end{align}
where the vector and axial coupling constants are defined in Eq.~\eqref{eq:OldGs}.

\subsection{Dirac Neutrinos}
After averaging over the initial-state photon polarization and electron spin and summing over final-state fermion spins, total matrix element squared $|\mathcal{M}|^2$ is comprised of the following elements:
\begin{equation}
\begin{aligned}
|\mathcal{M}_1|^2 &= \frac{64 e^2 G_F^2}{(s - m_e^2)^2} \times \\
&\left[\left(|g_A^{ i j}| + |g_V^{ i j}|\right)^2 (k_3 k_4) \left((k_5  p_1)s - m_e^2(3 k_5 p_1 + 2 k_5 p_2)\right)\right. \\
&\left. + \left(|g_A^{ i j}| - |g_V^{ i j}|\right)^2 (k_3 k_5) \left((k_4 p_1)s - m_e^2 (3 k_4 p_1 + 2k_4 p_2)\right)\right.\\
&\left. + \left(|g_V^{ i j}|^2 - |g_A^{ i j}|^2\right) (k_4 k_5) m_e^2 (s + m_e^2) \right].
\end{aligned}
\end{equation}
\begin{equation}
\begin{aligned}
|\mathcal{M}_2|^2 &= \frac{32 e^2 G_F^2}{(k_3 p_1)^2} \times \\
&\left[\left(|g_A^{ i j}| + |g_V^{ i j}|\right)^2 (k_5 p_2) \left( m_e^2 ((k_4 p_1) - (k_3 k_4)) + (k_3  p_1)(k_4  p_1)\right)\right. \\
&\left. + \left(|g_A^{ i j}| - |g_V^{ i j}|\right)^2 (k_4  p_2) \left( m_e^2 ((k_5  p_1) - (k_3  k_5)) + (k_3  p_1)(k_5  p_1)\right)\right. \\
&\left. + \left(|g_V^{ i j}|^2 - |g_A^{ i j}|^2\right) (k_4  k_5) m_e^2 (k_3  p_1 - m_e^2) \right]. 
\end{aligned}
\end{equation}
\begin{equation}
\begin{aligned}
&|\mathcal{M}_1^\dag \mathcal{M}_2 + \mathcal{M}_1 \mathcal{M}_2^\dag| = \frac{32 e^2 G_F^2}{(s - m_e^2)(k_3 p_1)} \times \\
&\left[\left(|g_A^{ i j}| + |g_V^{ i j}|\right)^2 \left((k_3 k_4) \left( ( k_3 k_5 + k_5 p_2)(m_e^2 - s) + 2 (k_3 p_1)(k_5 p_2) + 2 (k_3 p_2)((k_5 p_1) + 2(k_5 p_2))\right) \right.\right.\\
&\left.\left.+ 2 (k_3 p_1)(k_4 p_2)(k_5 p_2) - 2(k_3 p_2)(k_4 p_1)(k_5 p_2) \right) \right.\\
&\left. + \left(|g_A^{ i j}| - |g_V^{ i j}|\right)^2 \left((k_3 k_5) \left( ( k_3 k_4 + k_4 p_2)(m_e^2 - s) + 2 (k_3 p_1)(k_4 p_2) + 2 (k_3 p_2)((k_4 p_1) + 2(k_4 p_2))\right)\right.\right.\\
&\left.\left. + 2 (k_3 p_1)(k_4 p_2)(k_5 p_2) - 2(k_3 p_2)(k_4 p_2)(k_5 p_1)\right)\right.\\
&\left. + \left(|g_A^{ i j}|^2 - |g_V^{ i j}|^2\right) m_e^2 \left( (k_4 k_5)(2 (k_3 p_1) + 4(k_3 p_2) + m_e^2 - s) + 4 (k_4 p_1)(k_5 p_1) \right)\right],
\end{aligned}
\end{equation}
such that
\begin{equation}
|\mathcal{M}|^2 = |\mathcal{M}_1|^2 + |\mathcal{M}_2|^2 + |\mathcal{M}_1^\dag \mathcal{M}_2 + \mathcal{M}_1 \mathcal{M}_2^\dag|.
\end{equation}

\subsection{Majorana Neutrinos}
\label{subsec:MajAnalytic}
If neutrinos are Majorana in nature, then additional interference arises. For example, the Dirac final states $\nu_1 \overline{\nu}_2$ and $\nu_2 \overline{\nu}_1$ are distinguishable; for the Majorana case, they are not. Additional matrix elements from the interchange of $i$ and $j$ arise:
\begin{align}
i\mathcal{M}'_1 &= -\frac{e G_F}{\sqrt{2} (s - m_e^2)} \epsilon_\nu (p_1) \left[\overline{u}_{k_3} \gamma^\mu (g_V^{ j i} - g_A^{ j i}\gamma^5) (\slashed{p_1} + \slashed{p_2} + m_e) \gamma^\nu u_{p_2} \right] \left[ \overline{u}_{k_4} \gamma_\mu (1 + \gamma^5) v_{k_5} \right],\\
i\mathcal{M}'_2 &= -\frac{e G_F}{\sqrt{2} ((k_3 - p_1)^2 - m_e^2)} \epsilon_\nu (p_1) \left[ \overline{u}_{k_3} \gamma^\nu (\slashed{k_3} - \slashed{p_1} + m_e) \gamma^\mu (g_V^{ j i} - g_A^{ j i}\gamma^5) u_{p_2} \right] \left[ \overline{u}_{k_4} \gamma_\mu (1 + \gamma^5) v_{k_5} \right].
\end{align}

When $i = j$, the sum $i \mathcal{M} + i\mathcal{M}'$ simplifies:
\begin{align}
i(\mathcal{M}_1 + \mathcal{M}'_1) &= -2\frac{e G_F}{\sqrt{2} (s - m_e^2)} \epsilon_\nu (p_1) \left[\overline{u}_{k_3} \gamma^\mu (g_V^{ i i} - g_A^{ i i}\gamma^5) (\slashed{p_1} + \slashed{p_2} + m_e) \gamma^\nu u_{p_2} \right] \left[ \overline{u}_{k_4} \gamma_\mu \gamma^5 v_{k_5} \right],\\
i(\mathcal{M}_2 + \mathcal{M}'_2) &= -2\frac{e G_F}{\sqrt{2} ((k_3 - p_1)^2 - m_e^2)} \epsilon_\nu (p_1) \left[ \overline{u}_{k_3} \gamma^\nu (\slashed{k_3} - \slashed{p_1} + m_e) \gamma^\mu (g_V^{ i i} - g_A^{ i i}\gamma^5) u_{p_2} \right] \left[ \overline{u}_{k_4} \gamma_\mu \gamma^5 v_{k_5} \right].
\end{align}
In general, however, we must allow for the possibility of nontrivial Majorana phases $\alpha_i$. The most general way of writing the matrix elements, then, defining $\Delta \alpha_{ij} \equiv \alpha_i - \alpha_j$, is
\begin{align}
i(\mathcal{M}_1 + \mathcal{M}'_1) = 2\frac{e G_F}{\sqrt{2} (s - m_e^2)} \epsilon_\nu (p_1) &\left[\overline{u}_{k_3} \gamma^\mu (|g_V^{ i j}| - |g_A^{ i j}|\gamma^5) (\slashed{p_1} + \slashed{p_2} + m_e) \gamma^\nu u_{p_2} \right] \times \nonumber \\
&\left[ \overline{u}_{k_4} \gamma_\mu (i\sin{(\Delta \alpha_{ij})} + \cos{(\Delta \alpha_{ij})}\gamma^5) v_{k_5} \right],\\
i(\mathcal{M}_2 + \mathcal{M}'_2) = -\frac{e G_F}{\sqrt{2} ((k_3 - p_1)^2 - m_e^2)} \epsilon_\nu (p_1) &\left[ \overline{u}_{k_3} \gamma^\nu (\slashed{k_3} - \slashed{p_1} + m_e) \gamma^\mu (|g_V^{ i j}| - |g_A^{ i j}|\gamma^5) u_{p_2} \right] \times \nonumber \\
&\left[ \overline{u}_{k_4} \gamma_\mu (i\sin{(\Delta \alpha_{ij})} + \cos{(\Delta \alpha_{ij})}\gamma^5) v_{k_5} \right].
\end{align}

Squaring the matrix elements, we acquire terms proportional to $m_i m_j$, where $m_i$ is the mass of neutrino mass eigenstate $i$, that were not present in the Dirac case. Setting $m_i$ or $m_j = 0$ recovers the matrix-element-squared of the Dirac case, and corrections to this for the Majorana case are
\begin{equation}
\begin{aligned}
&\Delta |\mathcal{M}|^2_\text{Maj.} = \frac{32 e^2 G_F^2 m_i m_j \cos{(2\Delta \alpha_{ij})}}{(s-m_e^2)^2 (k_3 p_1)^2} \times \\
&\left[\left(|g_A^{ i j}|^2 + |g_V^{ i j}|^2\right) \left( 4 (k_3 p_1)^3 (3 m_e^2 -s) + 4 (k_3 p_1)^2 (4 (k_3 p_2) m_e^2 - 2(k_3 p_2) s + m_e^4 - m_e^2 s) \right.\right.\\
&\left.\left.+ (k_3 p_1)(m_e^2 - s)(8 (k_3 p_2)^2 - 4s((k_3 p_2) + m_e^2) + 4 (k_3 p_2) m_e^2 + 3 m_e^4 + s^2) + m_e^2 (s - m_e^2)^2 (2 (k_3 p_2) + 4 m_e^2 -s) \right)\right. \\
&\left. +4 m_e^2 \left(|g_A^{ i j}|^2 - |g_V^{ i j}|^2\right) \left( 4 (k_3 p_1)^2 m_e^2 + (s - m_e^2) \left( (s - m_e^2) m_e^2 - 4 (k_3 p_1)(k_3 p_2)\right) \right) \right].
\end{aligned}
\end{equation}

\section{Four-Neutrino Formalism}
\label{sec:Appendix2}
We revisit the four-point $\ell\ell\nu\nu$ interaction Eq.~\eqref{eq:interaction} in the context of nonzero mixing with a sterile neutrino. The Lagrangian of Eq.~\eqref{eq:WeakLagrangian} remains unchanged, but we emphasize that the implicit sum over $\alpha, \, \beta$ is only over the active flavor eigenstates, i.e., $e, \, \mu, \, \tau$. Using Eq.~\eqref{eq:DefineU} to replace the flavor eigenstates in the charged-current Lagrangian in favor of the mass eigenstates proceeds similarly to the three-neutrino case, the only difference being that the indices of $U_{\alpha i}$ can take the values $\alpha = e, \, \mu, \, \tau, \, s$ and $i = 1, \, 2, \, 3, \, 4$. In the neutral-current interaction, however, there is an additional subtlety: since the Lagrangian only sums over the active neutrino flavors, the Lagrangian will not end up diagonal after rotating to the neutrino mass basis. To wit, the neutrino neutral-current term becomes
\begin{equation}
\sum_{\alpha = e, \, \mu, \, \tau} \left( \overline{\nu}_\alpha \gamma^\mu P_L \nu_\alpha \right) = \sum_{i, \, j} \sum_{\alpha = e, \, \mu, \, \tau} U_{\alpha i} U_{\alpha j}^* \left( \overline{\nu}_j \gamma^\mu P_L \nu_i \right) = \sum_{i, \, j} \left(\delta_{ij} - U_{si} U_{sj}^*\right)  \left( \overline{\nu}_j \gamma^\mu P_L \nu_i \right);
\end{equation}
the result of this is to modify the vector and axial couplings $g_{V,A}^{ij} \left( \equiv g_{V,A}^{eeij} \right)$,
\begin{equation}
g_V^{ i j} = U_{e i} U_{e j}^* - \frac{1}{2}\left(1 - 4\sin^2{\theta_W}\right) \left(  \delta_{ij} - U_{si} U_{sj}^* \right), \qquad g_A^{ i j} = U_{e i} U_{e j}^* - \frac{1}{2} \left(  \delta_{ij} - U_{si} U_{sj}^* \right),
\end{equation}
precisely as stated in Eq.~\eqref{eq:NewGs}.


\bibliographystyle{apsrev-title}
\bibliography{eg_evv_bib}{}

\begin{thebibliography}{48}
\expandafter\ifx\csname natexlab\endcsname\relax\def\natexlab#1{#1}\fi
\expandafter\ifx\csname bibnamefont\endcsname\relax
  \def\bibnamefont#1{#1}\fi
\expandafter\ifx\csname bibfnamefont\endcsname\relax
  \def\bibfnamefont#1{#1}\fi
\expandafter\ifx\csname citenamefont\endcsname\relax
  \def\citenamefont#1{#1}\fi
\expandafter\ifx\csname url\endcsname\relax
  \def\url#1{\texttt{#1}}\fi
\expandafter\ifx\csname urlprefix\endcsname\relax\def\urlprefix{URL }\fi
\providecommand{\bibinfo}[2]{#2}
\providecommand{\eprint}[2][]{\url{#2}}

\bibitem[{\citenamefont{Esteban et~al.}(2017)\citenamefont{Esteban,
  Gonzalez-Garcia, Maltoni, Martinez-Soler, and Schwetz}}]{Esteban:2016qun}
\bibinfo{author}{\bibfnamefont{I.}~\bibnamefont{Esteban}},
  \bibinfo{author}{\bibfnamefont{M.~C.} \bibnamefont{Gonzalez-Garcia}},
  \bibinfo{author}{\bibfnamefont{M.}~\bibnamefont{Maltoni}},
  \bibinfo{author}{\bibfnamefont{I.}~\bibnamefont{Martinez-Soler}},
  \bibnamefont{and} \bibinfo{author}{\bibfnamefont{T.}~\bibnamefont{Schwetz}},
  ``{Updated fit to three neutrino mixing: exploring the accelerator-reactor
  complementarity},'' \bibinfo{journal}{JHEP} \textbf{\bibinfo{volume}{01}},
  \bibinfo{pages}{087} (\bibinfo{year}{2017}), \eprint{1611.01514}.

\bibitem[{\citenamefont{Kraus et~al.}(2005)}]{Kraus:2004zw}
\bibinfo{author}{\bibfnamefont{C.}~\bibnamefont{Kraus}} \bibnamefont{et~al.},
  ``{Final results from phase II of the Mainz neutrino mass search in tritium
  beta decay},'' \bibinfo{journal}{Eur. Phys. J.}
  \textbf{\bibinfo{volume}{C40}}, \bibinfo{pages}{447} (\bibinfo{year}{2005}),
  \eprint{hep-ex/0412056}.

\bibitem[{\citenamefont{Aseev et~al.}(2011)}]{Aseev:2011dq}
\bibinfo{author}{\bibfnamefont{V.~N.} \bibnamefont{Aseev}} \bibnamefont{et~al.}
  (\bibinfo{collaboration}{Troitsk}), ``{An upper limit on electron
  antineutrino mass from Troitsk experiment},'' \bibinfo{journal}{Phys. Rev.}
  \textbf{\bibinfo{volume}{D84}}, \bibinfo{pages}{112003}
  (\bibinfo{year}{2011}), \eprint{1108.5034}.

\bibitem[{\citenamefont{Osipowicz et~al.}(2001)}]{Osipowicz:2001sq}
\bibinfo{author}{\bibfnamefont{A.}~\bibnamefont{Osipowicz}}
  \bibnamefont{et~al.} (\bibinfo{collaboration}{KATRIN}), ``{KATRIN: A Next
  generation tritium beta decay experiment with sub-eV sensitivity for the
  electron neutrino mass. Letter of intent},''  (\bibinfo{year}{2001}),
  \eprint{hep-ex/0109033}.

\bibitem[{\citenamefont{Arenz et~al.}(2018)}]{Arenz:2018kma}
\bibinfo{author}{\bibfnamefont{M.}~\bibnamefont{Arenz}} \bibnamefont{et~al.}
  (\bibinfo{collaboration}{KATRIN}), ``{First transmission of electrons and
  ions through the KATRIN beamline},''  (\bibinfo{year}{2018}),
  \eprint{1802.04167}.

\bibitem[{\citenamefont{Ade et~al.}(2016)}]{Ade:2015xua}
\bibinfo{author}{\bibfnamefont{P.~A.~R.} \bibnamefont{Ade}}
  \bibnamefont{et~al.} (\bibinfo{collaboration}{Planck}), ``{Planck 2015
  results. XIII. Cosmological parameters},'' \bibinfo{journal}{Astron.
  Astrophys.} \textbf{\bibinfo{volume}{594}}, \bibinfo{pages}{A13}
  (\bibinfo{year}{2016}), \eprint{1502.01589}.

\bibitem[{\citenamefont{Dell'Oro et~al.}(2016)\citenamefont{Dell'Oro, Marcocci,
  Viel, and Vissani}}]{DellOro:2016tmg}
\bibinfo{author}{\bibfnamefont{S.}~\bibnamefont{Dell'Oro}},
  \bibinfo{author}{\bibfnamefont{S.}~\bibnamefont{Marcocci}},
  \bibinfo{author}{\bibfnamefont{M.}~\bibnamefont{Viel}}, \bibnamefont{and}
  \bibinfo{author}{\bibfnamefont{F.}~\bibnamefont{Vissani}}, ``{Neutrinoless
  double beta decay: 2015 review},'' \bibinfo{journal}{Adv. High Energy Phys.}
  \textbf{\bibinfo{volume}{2016}}, \bibinfo{pages}{2162659}
  (\bibinfo{year}{2016}), \eprint{1601.07512}.

\bibitem[{\citenamefont{Henning}(2016)}]{Henning:2016fad}
\bibinfo{author}{\bibfnamefont{R.}~\bibnamefont{Henning}}, ``{Current status of
  neutrinoless double-beta decay searches},'' \bibinfo{journal}{Rev. Phys.}
  \textbf{\bibinfo{volume}{1}}, \bibinfo{pages}{29} (\bibinfo{year}{2016}).

\bibitem[{\citenamefont{Atre et~al.}(2005)\citenamefont{Atre, Barger, and
  Han}}]{Atre:2005eb}
\bibinfo{author}{\bibfnamefont{A.}~\bibnamefont{Atre}},
  \bibinfo{author}{\bibfnamefont{V.}~\bibnamefont{Barger}}, \bibnamefont{and}
  \bibinfo{author}{\bibfnamefont{T.}~\bibnamefont{Han}}, ``{Upper bounds on
  lepton-number violating processes},'' \bibinfo{journal}{Phys. Rev.}
  \textbf{\bibinfo{volume}{D71}}, \bibinfo{pages}{113014}
  (\bibinfo{year}{2005}), \eprint{hep-ph/0502163}.

\bibitem[{\citenamefont{Berryman et~al.}(2017)\citenamefont{Berryman,
  de~Gouv\^{e}a, Kelly, and Kobach}}]{Berryman:2016slh}
\bibinfo{author}{\bibfnamefont{J.~M.} \bibnamefont{Berryman}},
  \bibinfo{author}{\bibfnamefont{A.}~\bibnamefont{de~Gouv\^{e}a}},
  \bibinfo{author}{\bibfnamefont{K.~J.} \bibnamefont{Kelly}}, \bibnamefont{and}
  \bibinfo{author}{\bibfnamefont{A.}~\bibnamefont{Kobach}},
  ``{Lepton-number-violating searches for muon to positron conversion},''
  \bibinfo{journal}{Phys. Rev.} \textbf{\bibinfo{volume}{D95}},
  \bibinfo{pages}{115010} (\bibinfo{year}{2017}), \eprint{1611.00032}.

\bibitem[{\citenamefont{Geib et~al.}(2017)\citenamefont{Geib, Merle, and
  Zuber}}]{Geib:2016atx}
\bibinfo{author}{\bibfnamefont{T.}~\bibnamefont{Geib}},
  \bibinfo{author}{\bibfnamefont{A.}~\bibnamefont{Merle}}, \bibnamefont{and}
  \bibinfo{author}{\bibfnamefont{K.}~\bibnamefont{Zuber}}, ``{$\mu^- - e^+$
  conversion in upcoming LFV experiments},'' \bibinfo{journal}{Phys. Lett.}
  \textbf{\bibinfo{volume}{B764}}, \bibinfo{pages}{157} (\bibinfo{year}{2017}),
  \eprint{1609.09088}.

\bibitem[{\citenamefont{Kuno}(2015)}]{Kuno:2015tya}
\bibinfo{author}{\bibfnamefont{Y.}~\bibnamefont{Kuno}}, ``{Rare lepton
  decays},'' \bibinfo{journal}{Prog. Part. Nucl. Phys.}
  \textbf{\bibinfo{volume}{82}}, \bibinfo{pages}{1} (\bibinfo{year}{2015}).

\bibitem[{\citenamefont{Quintero}(2017)}]{Quintero:2016iwi}
\bibinfo{author}{\bibfnamefont{N.}~\bibnamefont{Quintero}}, ``{Constraints on
  lepton number violating short-range interactions from $|\Delta L|=2$
  processes},'' \bibinfo{journal}{Phys. Lett.} \textbf{\bibinfo{volume}{B764}},
  \bibinfo{pages}{60} (\bibinfo{year}{2017}), \eprint{1606.03477}.

\bibitem[{\citenamefont{Akimov et~al.}(2017)}]{Akimov:2017ade}
\bibinfo{author}{\bibfnamefont{D.}~\bibnamefont{Akimov}} \bibnamefont{et~al.}
  (\bibinfo{collaboration}{COHERENT}), ``{Observation of Coherent Elastic
  Neutrino-Nucleus Scattering},'' \bibinfo{journal}{Science}
  \textbf{\bibinfo{volume}{357}}, \bibinfo{pages}{1123} (\bibinfo{year}{2017}),
  \eprint{1708.01294}.

\bibitem[{\citenamefont{Freedman}(1974)}]{Freedman:1973yd}
\bibinfo{author}{\bibfnamefont{D.~Z.} \bibnamefont{Freedman}}, ``{Coherent
  Neutrino Nucleus Scattering as a Probe of the Weak Neutral Current},''
  \bibinfo{journal}{Phys. Rev.} \textbf{\bibinfo{volume}{D9}},
  \bibinfo{pages}{1389} (\bibinfo{year}{1974}).

\bibitem[{\citenamefont{Erler and Su}(2013)}]{Erler:2013xha}
\bibinfo{author}{\bibfnamefont{J.}~\bibnamefont{Erler}} \bibnamefont{and}
  \bibinfo{author}{\bibfnamefont{S.}~\bibnamefont{Su}}, ``{The Weak Neutral
  Current},'' \bibinfo{journal}{Prog. Part. Nucl. Phys.}
  \textbf{\bibinfo{volume}{71}}, \bibinfo{pages}{119} (\bibinfo{year}{2013}),
  \eprint{1303.5522}.

\bibitem[{\citenamefont{Sabella et~al.}(2010)\citenamefont{Sabella, Piper, and
  Mildren}}]{Sabella:10}
\bibinfo{author}{\bibfnamefont{A.}~\bibnamefont{Sabella}},
  \bibinfo{author}{\bibfnamefont{J.~A.} \bibnamefont{Piper}}, \bibnamefont{and}
  \bibinfo{author}{\bibfnamefont{R.~P.} \bibnamefont{Mildren}}, ``1240 nm
  diamond raman laser operating near the quantum limit,''
  \bibinfo{journal}{Opt. Lett.} \textbf{\bibinfo{volume}{35}},
  \bibinfo{pages}{3874} (\bibinfo{year}{2010}),
  \urlprefix\url{http://ol.osa.org/abstract.cfm?URI=ol-35-23-3874}.

\bibitem[{\citenamefont{Peskin and Schroeder}(1995)}]{Peskin:1995ev}
\bibinfo{author}{\bibfnamefont{M.~E.} \bibnamefont{Peskin}} \bibnamefont{and}
  \bibinfo{author}{\bibfnamefont{D.~V.} \bibnamefont{Schroeder}},
  \emph{\bibinfo{title}{{An Introduction to quantum field theory}}}
  (\bibinfo{publisher}{Addison-Wesley}, \bibinfo{address}{Reading, USA},
  \bibinfo{year}{1995}), ISBN \bibinfo{isbn}{9780201503975, 0201503972},
  \urlprefix\url{http://www.slac.stanford.edu/~mpeskin/QFT.html}.

\bibitem[{\citenamefont{Ram and Wang}(1971)}]{Ram:1971pi}
\bibinfo{author}{\bibfnamefont{M.}~\bibnamefont{Ram}} \bibnamefont{and}
  \bibinfo{author}{\bibfnamefont{P.~Y.} \bibnamefont{Wang}}, ``{Calculation of
  the total cross-section for double Compton scattering},''
  \bibinfo{journal}{Phys. Rev. Lett.} \textbf{\bibinfo{volume}{26}},
  \bibinfo{pages}{476} (\bibinfo{year}{1971}).

\bibitem[{\citenamefont{Lotstedt and Jentschura}(2012)}]{Lotstedt:2012zz}
\bibinfo{author}{\bibfnamefont{E.}~\bibnamefont{Lotstedt}} \bibnamefont{and}
  \bibinfo{author}{\bibfnamefont{U.~D.} \bibnamefont{Jentschura}}, ``{Triple
  Compton Effect: A Photon Splitting into Three upon Collision with a Free
  Electron},'' \bibinfo{journal}{Phys. Rev. Lett.}
  \textbf{\bibinfo{volume}{108}}, \bibinfo{pages}{233201}
  (\bibinfo{year}{2012}), \eprint{1205.0317}.

\bibitem[{\citenamefont{Lotstedt and Jentschura}(2013)}]{Lotstedt:2013uya}
\bibinfo{author}{\bibfnamefont{E.}~\bibnamefont{Lotstedt}} \bibnamefont{and}
  \bibinfo{author}{\bibfnamefont{U.~D.} \bibnamefont{Jentschura}},
  ``{Theoretical study of the Compton effect with correlated three-photon
  emission: From the differential cross section to high-energy triple-photon
  entanglement},'' \bibinfo{journal}{Phys. Rev.}
  \textbf{\bibinfo{volume}{A87}}, \bibinfo{pages}{033401}
  (\bibinfo{year}{2013}), \eprint{1405.1669}.

\bibitem[{\citenamefont{Sudakov}(1956)}]{Sudakov:1954sw}
\bibinfo{author}{\bibfnamefont{V.~V.} \bibnamefont{Sudakov}}, ``{Vertex parts
  at very high-energies in quantum electrodynamics},'' \bibinfo{journal}{Sov.
  Phys. JETP} \textbf{\bibinfo{volume}{3}}, \bibinfo{pages}{65}
  (\bibinfo{year}{1956}), \bibinfo{note}{[Zh. Eksp. Teor. Fiz.30,87(1956)]}.

\bibitem[{\citenamefont{Yennie et~al.}(1961)\citenamefont{Yennie, Frautschi,
  and Suura}}]{Yennie:1961ad}
\bibinfo{author}{\bibfnamefont{D.~R.} \bibnamefont{Yennie}},
  \bibinfo{author}{\bibfnamefont{S.~C.} \bibnamefont{Frautschi}},
  \bibnamefont{and} \bibinfo{author}{\bibfnamefont{H.}~\bibnamefont{Suura}},
  ``{The infrared divergence phenomena and high-energy processes},''
  \bibinfo{journal}{Annals Phys.} \textbf{\bibinfo{volume}{13}},
  \bibinfo{pages}{379} (\bibinfo{year}{1961}).

\bibitem[{\citenamefont{J.~Gould}(1984)}]{GouldPaper}
\bibinfo{author}{\bibfnamefont{R.}~\bibnamefont{J.~Gould}}, ``The cross section
  for double compton scattering,'' \bibinfo{journal}{Astrophys. J.}
  \textbf{\bibinfo{volume}{285}}, \bibinfo{pages}{275} (\bibinfo{year}{1984}).

\bibitem[{\citenamefont{Dentler et~al.}(2018)\citenamefont{Dentler,
  Hern\'{a}sndez-Cabezudo, Kopp, Machado, Maltoni, Martinez-Soler, and
  Schwetz}}]{Dentler:2018sju}
\bibinfo{author}{\bibfnamefont{M.}~\bibnamefont{Dentler}},
  \bibinfo{author}{\bibfnamefont{A.}~\bibnamefont{Hern\'{a}sndez-Cabezudo}},
  \bibinfo{author}{\bibfnamefont{J.}~\bibnamefont{Kopp}},
  \bibinfo{author}{\bibfnamefont{P.}~\bibnamefont{Machado}},
  \bibinfo{author}{\bibfnamefont{M.}~\bibnamefont{Maltoni}},
  \bibinfo{author}{\bibfnamefont{I.}~\bibnamefont{Martinez-Soler}},
  \bibnamefont{and} \bibinfo{author}{\bibfnamefont{T.}~\bibnamefont{Schwetz}},
  ``{Updated global analysis of neutrino oscillations in the presence of
  eV-scale sterile neutrinos},''  (\bibinfo{year}{2018}), \eprint{1803.10661}.

\bibitem[{\citenamefont{Berryman et~al.}(2015)\citenamefont{Berryman,
  de~Gouv\^{e}a, Kelly, and Kobach}}]{Berryman:2015nua}
\bibinfo{author}{\bibfnamefont{J.~M.} \bibnamefont{Berryman}},
  \bibinfo{author}{\bibfnamefont{A.}~\bibnamefont{de~Gouv\^{e}a}},
  \bibinfo{author}{\bibfnamefont{K.~J.} \bibnamefont{Kelly}}, \bibnamefont{and}
  \bibinfo{author}{\bibfnamefont{A.}~\bibnamefont{Kobach}}, ``{Sterile neutrino
  at the Deep Underground Neutrino Experiment},'' \bibinfo{journal}{Phys. Rev.}
  \textbf{\bibinfo{volume}{D92}}, \bibinfo{pages}{073012}
  (\bibinfo{year}{2015}), \eprint{1507.03986}.

\bibitem[{\citenamefont{Henning et~al.}(2016)\citenamefont{Henning, Lu, and
  Murayama}}]{Henning:2014wua}
\bibinfo{author}{\bibfnamefont{B.}~\bibnamefont{Henning}},
  \bibinfo{author}{\bibfnamefont{X.}~\bibnamefont{Lu}}, \bibnamefont{and}
  \bibinfo{author}{\bibfnamefont{H.}~\bibnamefont{Murayama}}, ``{How to use the
  Standard Model effective field theory},'' \bibinfo{journal}{JHEP}
  \textbf{\bibinfo{volume}{01}}, \bibinfo{pages}{023} (\bibinfo{year}{2016}),
  \eprint{1412.1837}.

\bibitem[{\citenamefont{Brivio and Trott}(2017)}]{Brivio:2017vri}
\bibinfo{author}{\bibfnamefont{I.}~\bibnamefont{Brivio}} \bibnamefont{and}
  \bibinfo{author}{\bibfnamefont{M.}~\bibnamefont{Trott}}, ``{The Standard
  Model as an Effective Field Theory},''  (\bibinfo{year}{2017}),
  \eprint{1706.08945}.

\bibitem[{\citenamefont{Giunti and Studenikin}(2015)}]{Giunti:2014ixa}
\bibinfo{author}{\bibfnamefont{C.}~\bibnamefont{Giunti}} \bibnamefont{and}
  \bibinfo{author}{\bibfnamefont{A.}~\bibnamefont{Studenikin}}, ``{Neutrino
  electromagnetic interactions: a window to new physics},''
  \bibinfo{journal}{Rev. Mod. Phys.} \textbf{\bibinfo{volume}{87}},
  \bibinfo{pages}{531} (\bibinfo{year}{2015}), \eprint{1403.6344}.

\bibitem[{\citenamefont{Giunti et~al.}(2016)\citenamefont{Giunti, Kouzakov, Li,
  Lokhov, Studenikin, and Zhou}}]{Giunti:2015gga}
\bibinfo{author}{\bibfnamefont{C.}~\bibnamefont{Giunti}},
  \bibinfo{author}{\bibfnamefont{K.~A.} \bibnamefont{Kouzakov}},
  \bibinfo{author}{\bibfnamefont{Y.-F.} \bibnamefont{Li}},
  \bibinfo{author}{\bibfnamefont{A.~V.} \bibnamefont{Lokhov}},
  \bibinfo{author}{\bibfnamefont{A.~I.} \bibnamefont{Studenikin}},
  \bibnamefont{and} \bibinfo{author}{\bibfnamefont{S.}~\bibnamefont{Zhou}},
  ``{Electromagnetic neutrinos in laboratory experiments and astrophysics},''
  \bibinfo{journal}{Annalen Phys.} \textbf{\bibinfo{volume}{528}},
  \bibinfo{pages}{198} (\bibinfo{year}{2016}), \eprint{1506.05387}.

\bibitem[{\citenamefont{Balantekin and Kayser}(2018)}]{Balantekin:2018azf}
\bibinfo{author}{\bibfnamefont{A.~B.} \bibnamefont{Balantekin}}
  \bibnamefont{and} \bibinfo{author}{\bibfnamefont{B.}~\bibnamefont{Kayser}},
  ``{On the Properties of Neutrinos},''  (\bibinfo{year}{2018}),
  \eprint{1805.00922}.

\bibitem[{\citenamefont{Bressi et~al.}(2011)\citenamefont{Bressi, Carugno,
  Della~Valle, Galeazzi, Ruoso, and Sartori}}]{Bressi:2011yfa}
\bibinfo{author}{\bibfnamefont{G.}~\bibnamefont{Bressi}},
  \bibinfo{author}{\bibfnamefont{G.}~\bibnamefont{Carugno}},
  \bibinfo{author}{\bibfnamefont{F.}~\bibnamefont{Della~Valle}},
  \bibinfo{author}{\bibfnamefont{G.}~\bibnamefont{Galeazzi}},
  \bibinfo{author}{\bibfnamefont{G.}~\bibnamefont{Ruoso}}, \bibnamefont{and}
  \bibinfo{author}{\bibfnamefont{G.}~\bibnamefont{Sartori}}, ``{Testing the
  neutrality of matter by acoustic means in a spherical resonator},''
  \bibinfo{journal}{Phys. Rev.} \textbf{\bibinfo{volume}{A83}},
  \bibinfo{pages}{052101} (\bibinfo{year}{2011}), \eprint{1102.2766}.

\bibitem[{\citenamefont{Bernabeu et~al.}(2000)\citenamefont{Bernabeu,
  Cabral-Rosetti, Papavassiliou, and Vidal}}]{Bernabeu:2000hf}
\bibinfo{author}{\bibfnamefont{J.}~\bibnamefont{Bernabeu}},
  \bibinfo{author}{\bibfnamefont{L.~G.} \bibnamefont{Cabral-Rosetti}},
  \bibinfo{author}{\bibfnamefont{J.}~\bibnamefont{Papavassiliou}},
  \bibnamefont{and} \bibinfo{author}{\bibfnamefont{J.}~\bibnamefont{Vidal}},
  ``{On the charge radius of the neutrino},'' \bibinfo{journal}{Phys. Rev.}
  \textbf{\bibinfo{volume}{D62}}, \bibinfo{pages}{113012}
  (\bibinfo{year}{2000}), \eprint{hep-ph/0008114}.

\bibitem[{\citenamefont{Bernabeu et~al.}(2004)\citenamefont{Bernabeu,
  Papavassiliou, and Vidal}}]{Bernabeu:2002pd}
\bibinfo{author}{\bibfnamefont{J.}~\bibnamefont{Bernabeu}},
  \bibinfo{author}{\bibfnamefont{J.}~\bibnamefont{Papavassiliou}},
  \bibnamefont{and} \bibinfo{author}{\bibfnamefont{J.}~\bibnamefont{Vidal}},
  ``{The Neutrino charge radius is a physical observable},''
  \bibinfo{journal}{Nucl. Phys.} \textbf{\bibinfo{volume}{B680}},
  \bibinfo{pages}{450} (\bibinfo{year}{2004}), \eprint{hep-ph/0210055}.

\bibitem[{\citenamefont{Agostini et~al.}(2017)}]{Borexino:2017fbd}
\bibinfo{author}{\bibfnamefont{M.}~\bibnamefont{Agostini}} \bibnamefont{et~al.}
  (\bibinfo{collaboration}{Borexino}), ``{Limiting neutrino magnetic moments
  with Borexino Phase-II solar neutrino data},'' \bibinfo{journal}{Phys. Rev.}
  \textbf{\bibinfo{volume}{D96}}, \bibinfo{pages}{091103}
  (\bibinfo{year}{2017}), \eprint{1707.09355}.

\bibitem[{\citenamefont{Papoulias and Kosmas}(2018)}]{Kosmas:2017tsq}
\bibinfo{author}{\bibfnamefont{D.~K.} \bibnamefont{Papoulias}}
  \bibnamefont{and} \bibinfo{author}{\bibfnamefont{T.~S.}
  \bibnamefont{Kosmas}}, ``{COHERENT constraints to conventional and exotic
  neutrino physics},'' \bibinfo{journal}{Phys. Rev.}
  \textbf{\bibinfo{volume}{D97}}, \bibinfo{pages}{033003}
  (\bibinfo{year}{2018}), \eprint{1711.09773}.

\bibitem[{\citenamefont{Patrignani et~al.}(2016)}]{Patrignani:2016xqp}
\bibinfo{author}{\bibfnamefont{C.}~\bibnamefont{Patrignani}}
  \bibnamefont{et~al.} (\bibinfo{collaboration}{Particle Data Group}),
  ``{Review of Particle Physics},'' \bibinfo{journal}{Chin. Phys.}
  \textbf{\bibinfo{volume}{C40}}, \bibinfo{pages}{100001}
  (\bibinfo{year}{2016}).

\bibitem[{\citenamefont{Fujikawa and Shrock}(1980)}]{Fujikawa:1980yx}
\bibinfo{author}{\bibfnamefont{K.}~\bibnamefont{Fujikawa}} \bibnamefont{and}
  \bibinfo{author}{\bibfnamefont{R.}~\bibnamefont{Shrock}}, ``{The Magnetic
  Moment of a Massive Neutrino and Neutrino Spin Rotation},''
  \bibinfo{journal}{Phys. Rev. Lett.} \textbf{\bibinfo{volume}{45}},
  \bibinfo{pages}{963} (\bibinfo{year}{1980}).

\bibitem[{\citenamefont{Pal and Wolfenstein}(1982)}]{Pal:1981rm}
\bibinfo{author}{\bibfnamefont{P.~B.} \bibnamefont{Pal}} \bibnamefont{and}
  \bibinfo{author}{\bibfnamefont{L.}~\bibnamefont{Wolfenstein}}, ``{Radiative
  Decays of Massive Neutrinos},'' \bibinfo{journal}{Phys. Rev.}
  \textbf{\bibinfo{volume}{D25}}, \bibinfo{pages}{766} (\bibinfo{year}{1982}).

\bibitem[{\citenamefont{Shrock}(1982)}]{Shrock:1982sc}
\bibinfo{author}{\bibfnamefont{R.~E.} \bibnamefont{Shrock}}, ``{Electromagnetic
  Properties and Decays of Dirac and Majorana Neutrinos in a General Class of
  Gauge Theories},'' \bibinfo{journal}{Nucl. Phys.}
  \textbf{\bibinfo{volume}{B206}}, \bibinfo{pages}{359} (\bibinfo{year}{1982}).

\bibitem[{\citenamefont{Harnik et~al.}(2012)\citenamefont{Harnik, Kopp, and
  Machado}}]{Harnik:2012ni}
\bibinfo{author}{\bibfnamefont{R.}~\bibnamefont{Harnik}},
  \bibinfo{author}{\bibfnamefont{J.}~\bibnamefont{Kopp}}, \bibnamefont{and}
  \bibinfo{author}{\bibfnamefont{P.~A.~N.} \bibnamefont{Machado}}, ``{Exploring
  $\nu$ Signals in Dark Matter Detectors},'' \bibinfo{journal}{JCAP}
  \textbf{\bibinfo{volume}{1207}}, \bibinfo{pages}{026} (\bibinfo{year}{2012}),
  \eprint{1202.6073}.

\bibitem[{\citenamefont{Cerde\~{n}o et~al.}(2016)\citenamefont{Cerde\~{n}o,
  Fairbairn, Jubb, Machado, Vincent, and B{\oe}hm}}]{Cerdeno:2016sfi}
\bibinfo{author}{\bibfnamefont{D.~G.} \bibnamefont{Cerde\~{n}o}},
  \bibinfo{author}{\bibfnamefont{M.}~\bibnamefont{Fairbairn}},
  \bibinfo{author}{\bibfnamefont{T.}~\bibnamefont{Jubb}},
  \bibinfo{author}{\bibfnamefont{P.~A.~N.} \bibnamefont{Machado}},
  \bibinfo{author}{\bibfnamefont{A.~C.} \bibnamefont{Vincent}},
  \bibnamefont{and} \bibinfo{author}{\bibfnamefont{C.}~\bibnamefont{B{\oe}hm}},
  ``{Physics from solar neutrinos in dark matter direct detection
  experiments},'' \bibinfo{journal}{JHEP} \textbf{\bibinfo{volume}{05}},
  \bibinfo{pages}{118} (\bibinfo{year}{2016}), \bibinfo{note}{[Erratum:
  JHEP09,048(2016)]}, \eprint{1604.01025}.

\bibitem[{\citenamefont{Farzan et~al.}(2001)\citenamefont{Farzan, Peres, and
  Smirnov}}]{Farzan:2001cj}
\bibinfo{author}{\bibfnamefont{Y.}~\bibnamefont{Farzan}},
  \bibinfo{author}{\bibfnamefont{O.~L.~G.} \bibnamefont{Peres}},
  \bibnamefont{and} \bibinfo{author}{\bibfnamefont{A.~{\relax Yu}.}
  \bibnamefont{Smirnov}}, ``{Neutrino mass spectrum and future beta decay
  experiments},'' \bibinfo{journal}{Nucl. Phys.}
  \textbf{\bibinfo{volume}{B612}}, \bibinfo{pages}{59} (\bibinfo{year}{2001}),
  \eprint{hep-ph/0105105}.

\bibitem[{\citenamefont{Long et~al.}(2014)\citenamefont{Long, Lunardini, and
  Sabancilar}}]{Long:2014zva}
\bibinfo{author}{\bibfnamefont{A.~J.} \bibnamefont{Long}},
  \bibinfo{author}{\bibfnamefont{C.}~\bibnamefont{Lunardini}},
  \bibnamefont{and}
  \bibinfo{author}{\bibfnamefont{E.}~\bibnamefont{Sabancilar}}, ``{Detecting
  non-relativistic cosmic neutrinos by capture on tritium: phenomenology and
  physics potential},'' \bibinfo{journal}{JCAP}
  \textbf{\bibinfo{volume}{1408}}, \bibinfo{pages}{038} (\bibinfo{year}{2014}),
  \eprint{1405.7654}.

\bibitem[{\citenamefont{Vitagliano et~al.}(2017)\citenamefont{Vitagliano,
  Redondo, and Raffelt}}]{Vitagliano:2017odj}
\bibinfo{author}{\bibfnamefont{E.}~\bibnamefont{Vitagliano}},
  \bibinfo{author}{\bibfnamefont{J.}~\bibnamefont{Redondo}}, \bibnamefont{and}
  \bibinfo{author}{\bibfnamefont{G.}~\bibnamefont{Raffelt}}, ``{Solar neutrino
  flux at keV energies},'' \bibinfo{journal}{JCAP}
  \textbf{\bibinfo{volume}{1712}}, \bibinfo{pages}{010} (\bibinfo{year}{2017}),
  \eprint{1708.02248}.

\bibitem[{\citenamefont{Mertig et~al.}(1991)\citenamefont{Mertig, Bohm, and
  Denner}}]{Mertig:1990an}
\bibinfo{author}{\bibfnamefont{R.}~\bibnamefont{Mertig}},
  \bibinfo{author}{\bibfnamefont{M.}~\bibnamefont{Bohm}}, \bibnamefont{and}
  \bibinfo{author}{\bibfnamefont{A.}~\bibnamefont{Denner}}, ``{FEYN CALC:
  Computer algebraic calculation of Feynman amplitudes},''
  \bibinfo{journal}{Comput. Phys. Commun.} \textbf{\bibinfo{volume}{64}},
  \bibinfo{pages}{345} (\bibinfo{year}{1991}).

\bibitem[{\citenamefont{Shtabovenko et~al.}(2016)\citenamefont{Shtabovenko,
  Mertig, and Orellana}}]{Shtabovenko:2016sxi}
\bibinfo{author}{\bibfnamefont{V.}~\bibnamefont{Shtabovenko}},
  \bibinfo{author}{\bibfnamefont{R.}~\bibnamefont{Mertig}}, \bibnamefont{and}
  \bibinfo{author}{\bibfnamefont{F.}~\bibnamefont{Orellana}}, ``{New
  Developments in FeynCalc 9.0},'' \bibinfo{journal}{Comput. Phys. Commun.}
  \textbf{\bibinfo{volume}{207}}, \bibinfo{pages}{432} (\bibinfo{year}{2016}),
  \eprint{1601.01167}.

\bibitem[{\citenamefont{Ellis}(2017)}]{Ellis:2016jkw}
\bibinfo{author}{\bibfnamefont{J.}~\bibnamefont{Ellis}}, ``{TikZ-Feynman:
  Feynman diagrams with TikZ},'' \bibinfo{journal}{Comput. Phys. Commun.}
  \textbf{\bibinfo{volume}{210}}, \bibinfo{pages}{103} (\bibinfo{year}{2017}),
  \eprint{1601.05437}.

\end{thebibliography}

\end{document}